\documentclass[fleqn,usenatbib,useAMS]{mnras}

\usepackage{graphicx}	
\usepackage{amsmath}	
\usepackage{amssymb}	
\usepackage{multicol}        
\usepackage{bm}		
\usepackage{pdflscape}	

\newcommand{\gx}{\textsc{Gadget-X}}

\newcommand{\ahf}{\textsc{AHF}}

\newcommand{\Mbnd}{{\ifmmode{M_{\rm bnd}}\else{$M_{\rm bnd}$}\fi}}
\newcommand{\Mfof}{{\ifmmode{M_{\rm fof}}\else{$M_{\rm fof}$}\fi}}
\newcommand{\Mcrit}{{\ifmmode{M_{\rm 200c}}\else{$M_{\rm 200c}$}\fi}}
\newcommand{\Rcrit}{{\ifmmode{R_{\rm 200c}}\else{$R_{\rm 200c}$}\fi}}
\newcommand{\Rhost}{{\ifmmode{R_{\rm host}}\else{$R_{\rm host}$}\fi}}
\newcommand{\Mmean}{{\ifmmode{M_{\rm 200m}}\else{$M_{\rm 200m}$}\fi}}
\newcommand{\MBN}{{\ifmmode{M_{\rm BN98}}\else{$M_{\rm BN98}$}\fi}}

\newcommand{\hGpc}{{\ifmmode{h^{-1}{\rm Gpc}}\else{$h^{-1}$Gpc}\fi}}
\newcommand{\hMpc}{{\ifmmode{h^{-1}{\rm Mpc}}\else{$h^{-1}$Mpc}\fi}}
\newcommand{\hkpc}{{\ifmmode{h^{-1}{\rm kpc}}\else{$h^{-1}$kpc}\fi}}
\newcommand{\hMsun}{{\ifmmode{h^{-1}{\rm {M_{\odot}}}}\else{$h^{-1}{\rm{M_{\odot}}}$}\fi}}
\newcommand{\Mstar}{{\ifmmode{M_{*}}\else{$M_{*}$}\fi}}
\newcommand{\Mhalo}{{\ifmmode{M_{\rm Halo}}\else{$M_{\rm Halo}$}\fi}}
\newcommand{\Ngal}{{\ifmmode{N_{\rm gal}}\else{$N_{\rm gal}$}\fi}}
\newcommand{\Norph}{{\ifmmode{N_{\rm orphan}}\else{$N_{\rm orphan}$}\fi}}
\newcommand{\Nxorph}{{\ifmmode{N_{\rm non-orphan}}\else{$N_{\rm non-orphan}$}\fi}}
\newcommand{\Zsolar}{{\ifmmode{Z_{\odot}}\else{$Z_{\odot}$}\fi}}
\newcommand{\Msun}{{\ifmmode{{\rm {M_{\odot}}}}\else{${\rm{M_{\odot}}}$}\fi}}
\newcommand{\ltsima}{$\; \buildrel < \over \sim \;$}
\newcommand{\gtsima}{$\; \buildrel > \over \sim \;$}
\newcommand{\lsim}{\lower.5ex\hbox{\ltsima}}
\newcommand{\gsim}{\lower.5ex\hbox{\gtsima}}

\newcommand{\Tab}[1]{Table~\ref{#1}}
\newcommand{\Sec}[1]{Section~\ref{#1}}

\newcommand{\Eq}[1]{Eq.~(\ref{#1})}
\newcommand{\Fig}[1]{Fig.~\ref{#1}}
\newcommand{\beq}{\begin{equation}}
\newcommand{\eeq}{\end{equation}}

\usepackage[T1]{fontenc}
\usepackage{ae,aecompl}

\usepackage{newtxtext,newtxmath}

\title[Galaxy cluster mergers]{The Three Hundred project: galaxy cluster mergers and their impact on the stellar component of brightest cluster galaxies}

\author[Contreras-Santos et al.]
{Ana Contreras-Santos,$^{1}$\thanks{Contact e-mail: \href{mailto:ana.contreras@uam.es}{ana.contreras@uam.es}}
Alexander Knebe,$^{1,2,3}$
Frazer Pearce,$^{4}$
Roan Haggar,$^{4}$
\newauthor
Meghan Gray,$^{4}$
Weiguang Cui,$^{5}$
Gustavo Yepes,$^{1,2}$
Marco De Petris,$^{6,7}$
Federico De Luca,$^{8}$
\newauthor
Chris Power,$^{3}$
Robert Mostoghiu,$^{1}$
Sebasti\'an~E.~Nuza,$^{9,10}$
and Matthias Hoeft$^{11}$
\\
$^{1}$Departamento de F\'isica Te\'{o}rica, M\'{o}dulo 15, Facultad de Ciencias, Universidad Aut\'{o}noma de Madrid, 28049 Madrid, Spain\\
$^{2}$Centro de Investigaci\'{o}n Avanzada en F\'isica Fundamental (CIAFF), Facultad de Ciencias, Universidad Aut\'{o}noma de Madrid, 28049 Madrid, Spain\\
$^{3}$International Centre for Radio Astronomy Research, University of Western Australia, 35 Stirling Highway, Crawley, Western Australia 6009, Australia\\
$^{4}$School of Physics \& Astronomy, University of Nottingham, Nottingham NG7 2RD, United Kingdom\\
$^{5}$Institute for Astronomy, University of Edinburgh, Royal Observatory, Edinburgh EH9 3HJ, United Kingdom\\
$^{6}$Dipartimento di Fisica, Sapienza Università di Roma, Piazzale Aldo Moro 5, 00185 Roma, Italy\\
$^{7}$I.N.A.F. - Osservatorio Astronomico di Roma, Via Frascati 33, 00040 Monteporzio Catone, Roma, Italy\\
$^{8}$Dipartimento di Fisica, Università di Roma “Tor Vergata”, Via della Ricerca Scientifica 1, 00133 Roma, Italy\\
$^{9}$Instituto de Astronom\'{\i}a y F\'{\i}sica del Espacio (IAFE, CONICET-UBA), CC 67, Suc. 28, 1428 Buenos Aires, Argentina\\
$^{10}$Facultad de Ciencias Exactas y Naturales (FCEyN), Universidad de Buenos Aires (UBA), Buenos Aires, Argentina\\
$^{11}$Th\"{u}ringer Landessternwarte, Sternwarte 5, 07778 Tautenburg, Germany\\
}

\date{Last updated 2015 May 22; in original form 2013 September 5}

\pubyear{2021}

\begin{document}
\label{firstpage}
\pagerange{\pageref{firstpage}--\pageref{lastpage}}
\maketitle

\begin{abstract}
 Using the data set of \textsc{The Three Hundred} project, i.e. a suite of 324 hydrodynamical resimulations of cluster-sized haloes, we study galaxy cluster mergers and their effect on colour and luminosity changes of their brightest cluster galaxies (BCG). We track the main progenitor of each halo at $z=0$ and search for merger situations based on its mass accretion history, defining mergers as very rapid increases in the halo mass. Based upon the evolution of the dynamical state of the cluster we define a pre- and post-merger phase. We create a list of all these events and statistically study their mass ratio and timescales, with the former verifying that all instances are in fact major mergers. By comparing to a control sample of clusters without mergers, we study the effect mergers have on the stellar component of the BCG. Analysing the mass, age and metallicity of the BCG stellar particles, we find that the stellar content of BCGs grows significantly during mergers and, even though the main growth mechanism is the accretion of older stars, there is even a burst in star formation induced by the merger. In our simulations, BCGs in mergers form in median around 70 per cent more stars than those normally growing, although this depends on the radius considered for defining the BCG. Regarding observable properties, we see an increase in SDSS-$u$ luminosity of 20 per cent during mergers, accompanied by a slightly slower increase of the galaxy $g-r$ colour as compared to the control sample.
\end{abstract}

\begin{keywords}
  methods: numerical -- galaxies: clusters: general -- galaxies: haloes -- cosmology: theory -- large-scale structure of the universe 
\end{keywords}



\section{Introduction}

Galaxy clusters are essential in our comprehension of the Universe. They are the largest gravitationally bound systems and, as such, they can be used to probe the large scale structure of the Universe, as well as the formation and evolution of galaxies. On cosmological scales, they are dark matter dominated, and so their physics are only driven by gravity. On smaller scales, not only gravity is taken into account  but also the interaction of the baryonic components of clusters plays an important role, leading to several different phenomena that regulate for instance the properties of the hot gas in the intracluster medium (ICM). The study of galaxy clusters can therefore yield results ranging from cosmological parameters to models of the astrophysical processes that drive galaxy evolution.

Regarding the formation of galaxy clusters, the $\Lambda$ cold dark matter ($\Lambda$CDM) model of the Universe describes a hierarchical model of structure formation \citep{Blumenthal84}. Small bound structures are formed via gravitational collapse, which then grow through mergers with other haloes or via accretion of smaller systems (\citealp{White78,Frenk12}). This way, large haloes are a result of merger processes throughout their history and, thus, the understanding of these events becomes crucial in understanding the formation of clusters and their properties. Besides, due to the high binding energy and the huge energy releases, cluster mergers are one of the most energetic events in the Universe, which also makes them a very relevant field of research.

Several investigations have been conducted in this field both observationally and theoretically. Multiwavelength observations of galaxy clusters can probe differently the underlying physics. Observations in radio and far infrared probe the cold gas \citep{Giard2008}, while the optical (and near infrared) emission comes from stars, with two main observables being the luminosity and color (see \citealp{Bahcall77} or \citealp{Biviano2000} for reviews of optical studies of galaxy clusters). X-ray observations probe the hot intracluster medium (ICM), which shines brightly at these wavelengths (see \citealp{Sarazin1988} or \citealp{Bohringer2010} for reviews of X-ray observations of clusters).
The ICM is also explored by the Sunyaev-Zeldovich (SZ) signal at millimetre wavelengths, reaching even larger radii and distant clusters (see review by \citealp{Mroczkowski2019}). 

With these different observations, cluster surveys can be constructed, see e.g. for X-ray: ROSAT All-Sky Survey RASS, \citep{Voges1996}, CHEX-MATE, \citep{Chex-Mate2020}, eROSITA, \citep{eRosita2021};  for SZ: Planck \citep{Planck2015SZ}, ACT \citep{Hilton2018}, SPT \citep{Bleem2020}; or \citealp{Wen2009} for optical. Then, via their morphology, clusters undergoing mergers can be identified  \citep{Schombert1987,Mann2012}. Observations of merging clusters allow then a more in-depth study of these events \citep{Belsole2005,Okabe2008,Golovich2019} and the different phases in which they can be observed \citep{Wilber2019}. They also facilitate the analysis of the relationship between mergers and different cluster properties, such as kinetic energy and entropy \citep{Markevitch99} or star formation rates for the individual galaxies in the cluster \citep{Johnston-Hollitt2008,Deshev2017}. Other studies investigate the influence of mergers on the presence or absence of cool cores in clusters and on the different scaling relations like the X-ray luminosity-temperature relation \citep{Hallman2004,Ohara2006,Wang_Markevitch2016}.

Numerical simulations have also been used to study galaxy clusters from a more theoretical approach (see \citealp{Borgani2011} for a review), and their properties can be compared to those from observations \citep{Borgani2004,Fabjan2011}. 
Regarding galaxy cluster mergers, they can be studied in controlled simulations, where only two clusters are simulated to merge, so that the initial conditions and outcomes can be thoroughly studied \citep{Poole2006,Poole2007,Valdarnini2021}. This allows for a very detailed study, with the possibility to change the initial conditions as desired. However, it can be of even more interest to study mergers in the frame of cosmological simulations, where they happen naturally during the evolution of clusters (e.g. the Millennium Simulation, \citealp{springel_millennium_2005}). Unlike observations, this kind of simulations allow for tracking the whole history of a cluster, identifying a merger the moment it takes place. They are also useful to study the merger rate of dark matter haloes in the universe, and find its dependence with redshift or mass of the haloes \citep{Fakhouri08,Fakhouri10}. 
As with observations, data from simulations can be used to study the effects that mergers have on the already mentioned scaling relations or presence of cool cores \citep{Ritchie2002,Kay2007,Planelles2009,ZuHone2011}, and on different cluster properties such as their magnetic field \citep{Roettiger99,Brzycki2019}, halo shape and spin \citep{Vitvitska02,Moore2004,McMillan2007,Drakos2019a}, DM density profiles \citep{Kazantzidis2006,Drakos2019b} or the alignment between the DM halo and the central galaxy \citep{Ragone-Figueroa2020}.

In this paper we focus on the impact that mergers have on the stellar component of clusters and, particularly, in the stellar component of the brightest cluster galaxy (BCG). BCGs include the most massive and luminous galaxies in the Universe. They are large and elliptical galaxies, generally located right at the centre of their host cluster. Consequently, their formation and evolution are closely linked to those of the cluster, and hence different from typical elliptical galaxies \citep{LinMohr2004,Brough2005}. In the hierarchical formation scenario, theoretical models have predicted BCGs to assemble most of their mass through dry mergers with other galaxies \citep{Dubinski1998,Ruszkowski2009}. Regarding the stellar component of BCGs, recent studies -- based on full physics hydrodynamical simulations -- have investigated BCG growth and compared to observational data, finding good correspondence. They find that the mass growth was moderate since $z=1$, with a growth factor $\lesssim 2$ in this period \citep{Martizzi2016,Ragone-Figueroa2018}. 
Other studies, based on semi-analytical models (SAMs), show that the greatest part of star formation in BCGs took place before $z \sim 2$ \citep{DeLucia06,delucia_sam_2007}. In this sense, most BCGs in the observed universe are passive \citep{FraserMckelvie2014}, with star formation occurring mostly at high redshifts. However, it is also known that some BCGs have a significant amount of star formation at low redshifts. Although this is in general related to clusters with cool cores \citep{Donahue2010,Liu2012}, this is not a closed topic and the fraction of BCGs with star formation is unclear \citep{Runge2018}.

In this work we use `The Three Hundred' data set, that consists of regions of diameter 30$\hMpc$ centred on the 324 most massive objects found within a cosmological dark matter only simulation of side length 1$\hGpc$. Those regions have been re-simulated with \gx, i.e. full physics hydrodynamical code for cosmological simulations based upon a modern SPH (Smoothed-Particle Hydrodynamics) solver (see \citealp{Cui18} for more details about the code and data set). Using these simulations, we track the central object in each region from $z=0$ up to the highest redshift where it can be found. By looking at the mass of each object and its progenitors we find merger events and then study the effect they have on the involved clusters. To study how the stellar component of the BCGs is affected by cluster mergers we will first analyse directly the stellar particles in our simulations. Then, we will also study the luminosity and colour of these galaxies, to assess how mergers influence observations of BCGs.

The paper is organized as follows. In \Sec{sec:data}, we present the details of the simulation and the halo catalogues and merger trees used to track the haloes. In \Sec{sec:mergers} we present the definition used to find mergers and how to define their duration within our cluster sample. We present the mergers found and describe them in terms of some of their properties. In \Sec{sec:results} we analyse the effects of cluster mergers on their stellar component by comparing a merger sample and a control sample. Finally, in \Sec{sec:conclusions}, we summarize and discuss our results.

\section{The Data} \label{sec:data}

\subsection{The Three Hundred Clusters}
The 324 clusters in \textsc{The Three Hundred} data set were created upon the DM-only MDPL2 MultiDark Simulation \citep{Klypin16}, which is a periodic cube of comoving length 1 $\hGpc$ containing $3840^3$ DM particles, each of mass $1.5\cdot 10^9$ $\hMsun$. The Plummer equivalent softening of this simulation is 6.5 $\hkpc$. The cosmological parameters of the MDPL2 simulation are based on the Planck 2015 cosmology \citep{Planck2015}. From this simulation, the 324 clusters with the largest halo virial mass\footnote{The halo virial mass is defined as the mass enclosed inside an overdensity of $\sim$98 times the critical density of the universe \citep{Bryan98}} at $z=0$ with $M_\mathrm{vir} \gtrapprox 8 \cdot 10^{14} \hMsun$ were selected. These clusters serve as the centre of spherical regions with radius 15 $\hMpc$, where the initial DM particles were split into dark matter and gas particles (with masses $m_\mathrm{DM}=1.27 \times 10^9$ $\hMsun$ and $m_\mathrm{gas}=2.36 \times 10^8$ $\hMsun$ respectively), according to the cosmological baryon fraction. These regions were then re-simulated from their initial conditions including now full hydrodynamics using the SPH code \gx. Lower-resolution particles were used beyond 15 $\hMpc$, to replicate any large-scale tidal effects on the cluster at a lower computational cost. The output includes, for each of the 324 clusters, 129 snapshots between $z=0$ and $z=16.98$. At $z=0$, the 324 galaxy clusters have a mass range from $M_{200} = 6.4 \cdot 10^{14} \hMsun$ to $M_{200} = 2.65 \cdot 10^{15} \hMsun$. The size of our sample is such that it allows for statistically significant subsamples to be constructed. \textsc{The Three Hundred} data set was presented in an introductory paper by \citet{Cui18}, and several other papers have been published based on this data (see e.g. \citealp{Wang18,Mostoghiu18,Haggar2020,Herbonnet2021}), to which we refer the reader for more details about this project.

Regarding the code used for the re-simulations, \textsc{Gadget-X} is a modified version of the non-public \textsc{Gadget3} code \citep{Murante2010,Rasia2015,Planelles2017,Biffi2017}, which evolves dark matter with the \textsc{Gadget3} Tree-PM gravity solver (an advanced version of the \textsc{Gadget2} code; \citealp{springel_gadget2_2005}). It uses an improved SPH scheme that includes artificial thermal diffusion, time-dependent artificial viscosity, high-order Wendland C4 interpolating kernel and wake-up scheme (see \citealp{Beck2016} and \citealp{Sembolini2016} for a presentation of the performance of this SPH algorithm). Star formation is carried out as in \citet{Tornatore2007}, and follows the star formation algorithm presented in \citet{Springel03}. Black hole (BH) growth and AGN feedback are implemented following \citet{Steinborn2015}, where super massive black holes (SMBHs) grow via Bondi-Hoyle like gas accretion (Eddington limited), with the model distinguishing between a cold and a hot component.

\subsection{The Halo Catalogues \& Merger Trees}
All data were analysed with the open-source AHF halo finder \citep{Gill04a,Knollmann09}, which includes both gas and stars in the halo finding process. Haloes, as well as substructures, are found by locating overdensities in an adaptively smoothed density field (see e.g. \citealp{Knebe11} for more details on halo finders). For each halo identified, AHF computes its $R_{200}$ radius, which is the radius $r$ at which the density $\rho(r)=M(<r)/(4\pi r^3/3)$ drops below $200\rho_{\mathrm{crit}}$, where $\rho_{\mathrm{crit}}$ is the critical density of the Universe at the respective redshift. $R_{500}$, as well as the corresponding enclosed masses $M_{200}$ and $M_{500}$, is defined accordingly. Subhaloes are defined as haloes which lie within the $R_{200}$ region of a more massive halo, the so-called host halo.

The luminosity (and magnitude) in any spectral band from the stars within the haloes is calculated by applying the stellar population synthesis code \textsc{stardust} (see \citealp{Devriendt99}, and references therein for more details). This code computes the spectral energy distribution (SED) from far-UV to radio, for an instantaneous starburst of a given mass, age, and metallicity. The stellar contribution to the total flux is calculated assuming a Kennicutt initial mass function \citep{Kennicutt98}.

Finally, in order to follow the evolution of haloes with redshift, we need to trace them through the different snapshots. Merger trees are built for this purpose with \textsc{MergerTree}, a tool that comes with the AHF package. \textsc{MergerTree} follows each halo identified at $z=0$ backwards in time, using a merit function to identify the main progenitor, as well as other progenitors, in the previous snapshot. This tool also allows for skipping snapshots, so that the halo merger tree does not have to be truncated if no suitable progenitor is found in the immediately preceding snapshot. The main progenitor of halo A is the halo B (at a previous redshift) that maximizes the merit function: $\mathcal{M}=N^2_{AB}/(N_A N_B)$, where $N_A$ and $N_B$ are the number of particles in haloes A and B respectively, and $N_{AB}$ is the number of particles that are in both haloes A and B. For more details on the performance of \textsc{MergerTree} (and also different treebuilders) see \citet{Srisawat13}.

\section{Galaxy Cluster Mergers} \label{sec:mergers}

In contrast to binary merger simulations, where all the focus is on the two merging clusters (e.g.  \citealp{Poole2006,Poole2007,Donnert2013}), cosmological simulations require a clear method to find mergers and distinguish them from other events in the evolution of a halo (e.g. \citealp{Yu2015,Nuza2012,Nuza2017}). Simulations also allow to determine with clarity different merger properties, that can then be compared against other works or even observations. These properties include the mass ratio between the two merging clusters, which can significantly influence the outcome of a merger (see \citealp{Ricker2001,ZuHone2011}), and the length of the whole merger event. Detailed studies of the relaxation process after mergers can be found in \citealp{Valluri07} (N-body simulations of controlled mergers) and \citealp{Faltenbacher2006} (study of one merging event at $z=0.6$ from a high resolution cosmological N-body simulation).
 
In this Section we address the question of how to describe and study mergers in \textsc{The Three Hundred} dataset. We start by defining mergers and how to find them, and then classify them based on the duration of their effects in the respective cluster.

\subsection{Merger Definition: Mass Accretion History} \label{sec:definition}
The mass accretion history (MAH) of a cluster is obtained from the merger trees, which track haloes from $z=0$ up to the highest redshift where their progenitors are found (always smaller than the maximum redshift in the snapshots, $z=16.98$).
In this work, we are going to focus only on the central halo at $z=0$ of each of the 324 regions and, initially and for defining mergers, only on the main branch of its MAH. Similarly to other works (see \citealp{Wetzel2007} and \citealp{Cohn2005}), we can define a merger as a very rapid increase in the mass of a halo, as opposed to a slow accretion of many small objects over a long period of time. This way, in order to find mergers in our simulated data, we need a clear definition of this `rapid' increase. The previously mentioned works simply compare the mass of a halo in two consecutive snapshots, and require a minimum increase in the mass (e.g. $25-50$ per cent). However, the time elapsed between two snapshots can change with redshift and might not have any relation to timescales of physical processes. We therefore see this decision as somewhat arbitrary. For this reason, to set a time in which we can look for significant mass increases we are going to use the dynamical time of clusters, $t_d$. The most common way to define this (see e.g. \citealp{bt}) is the crossing time, i.e., the time it takes for a particle to complete a significant fraction of its orbit, which can be written as:
\beq
t_d \simeq \frac{R_{200}}{V_\mathrm{circ}},
\label{eq:dyntime}
\eeq
where $V_\mathrm{circ}$ is the circular velocity, obtained as $\sqrt{(GM_{200})/R_{200}}$. 
Given that the mass can be written in terms of the critical density $\rho_\mathrm{crit}$ as $M_{200}= 200 \rho_\mathrm{crit}(z) \frac{4\pi}{3} R_\mathrm{200}^3$, the dynamical time can also be written as:
\beq
t_d = \sqrt{\frac{3}{4\pi}\frac{1}{200 G \rho_\mathrm{crit}}}.
\label{eq:dyntime2}
\eeq

And since the critical density depends only on the cosmology as 
$\rho_\mathrm{crit}(z)=3/ (8 \pi G) \cdot H_0^2 (\Omega_{M,0}(1+z)^{3}+\Omega_{\Lambda,0})$ we see that the dynamical time evolves with redshift, but it is the same for all the clusters, regardless of their mass or radius. 

 Since we are interested in significant mass increases over a (short enough) period of time, we are going to use half of this dynamical time, looking for an increase in the mass of 100 per cent during that period. 
In terms of the mass of the cluster, this can also be written as $M_f \geq 2 M_i$, where $M_i$ ($M_f$) is the mass of the cluster at time $t_i$ ($t_f$) with the difference between those times obeying $t_f-t_i\leq t_d/2$. In \textsc{The Three Hundred} data set, with its chosen spacing between snapshots, we find that for all redshifts, two snapshots correspond to $ \lesssim t_d/2$. Using the fractional mass change 
\begin{equation}
\frac{\Delta M}{M} = \frac{M_f-M_i}{M_i}, 
\label{eq:deltaM/M}
\end{equation}
our condition for a merger (i.e. $M_f/M_i \geq 2$) translates into $\Delta M/M \geq 1$. We will study the evolution of $\Delta M/M$, using it to identify all the mergers a cluster undergoes, but also the periods of slower accretion. We further like to remark that using two snapshots is an upper limit: if the required mass increase is reached in only one snapshot, then we also identify that as a merger. Besides, it can also be the case that after identifying all the mergers in the MAH of a cluster, we end up with consecutive mergers, with one starting in the snapshot right after the one where the previous merger ended. We do not consider these to be two different mergers and hence combine them into a single one.

In summary, our definition of mergers consists of finding a mass increase of 100 per cent happening within half the cluster's dynamical time, allowing for possible extension over longer periods in case a `new' merger was found right afterwards. We search for such instances starting at the highest available redshift for the MAH of the central halo of each region, and iterating forward until $z=0$, obtaining a list of situations classified as mergers, with their respective redshifts and mass increase ratios.
Other works, like \citealp{Planelles2009} and \citealp{ChenImprints2019}, additionally compare the masses of the two main progenitors, requiring for a merger that they are comparable. We will return to this mass ratio later, using it instead as a way to characterise mergers as opposed to adding it to its definition.

It is also worth mentioning that, when studying the mass evolution of haloes, we have to be aware of their pseudo-evolution (see \citealp{Diemer13}). Given that the definition of $M_{200}$ is based on the critical density $\rho_\mathrm{crit}$, there is an evolution of halo mass only due to this reference density evolving with time, even if there is no physical accretion. This is called pseudo-evolution, and in \citet{Diemer13} it is shown that it can account for a very significant amount of the mass growth. However, this effect is specially important for galaxy-sized haloes ($M_{200} \lesssim 10^{12} \Msun$), and not for cluster-sized haloes like the ones we are working with ($M_{200} > 10^{14} \Msun$). Besides, pseudo-evolution implies a slower and more continuous mass growth than what we are defining as mergers, which involve short but significant mass increases. In this sense, given our required $\Delta M/M \geq 1$ threshold, we can be confident that the mass increases found are not due to pseudo-evolution but to physical merger events.

\subsection{Dynamical State Evolution}

Mergers are not only associated with a mass growth but also with a disturbance of the clusters' dynamical state. Since mergers are fast increases in the mass of a cluster, they are expected to significantly disturb clusters. In turn, after enough time has elapsed and these effects weaken, clusters should be dynamically relaxed again. For this reason, we are also going to study how the dynamical state of the clusters in our sample evolves with time, so that we can thus use its evolution around mergers to define pre- and post-merger phases.

To quantify the dynamical state of clusters we use the so-called `relaxation parameter', introduced by \citet{Haggar2020} as:
\begin{equation}
\chi_{\mathrm{DS}}=\sqrt{\frac{3}{\left(\frac{\Delta_r}{0.04}\right)^2+\left(\frac{f_s}{0.1}\right)^2+\left(\frac{\vert 1 - \eta \vert}{0.15}\right)^2}}.
\label{eq:chiDS}
\end{equation}
This equation is based on the three parameters initially introduced by \citet{Neto07} as proxies for relaxation, which were later used by \citet{Cui18} to study the relaxation of the clusters in \textsc{The Three Hundred} sample (another study by \citealp{DeLuca2021} uses only the first two parameters). The parameters are: 
\begin{itemize}
    \item \textbf{centre of mass offset}, \bm{$\Delta_r$}: offset of the centre of mass of the cluster from the density peak of the cluster halo, as a fraction of the cluster radius $R_{200}$;
    \item \textbf{subhalo mass fraction}, \bm{$f_s$}: fraction of the cluster mass contained in subhaloes; and
    \item \textbf{virial ratio}, \bm{$\eta$}, defined as $\eta=(2T-E_s)/ \vert W \vert$, where $T$ is the total kinetic energy of the cluster, $E_s$ its energy from surface pressure for both gas and collision-less particles \citep{Cui2017} and $W$ its total potential energy.
\end{itemize}

\begin{figure*}
   \hspace*{-0.1cm}
   {\includegraphics[width=8.5cm]{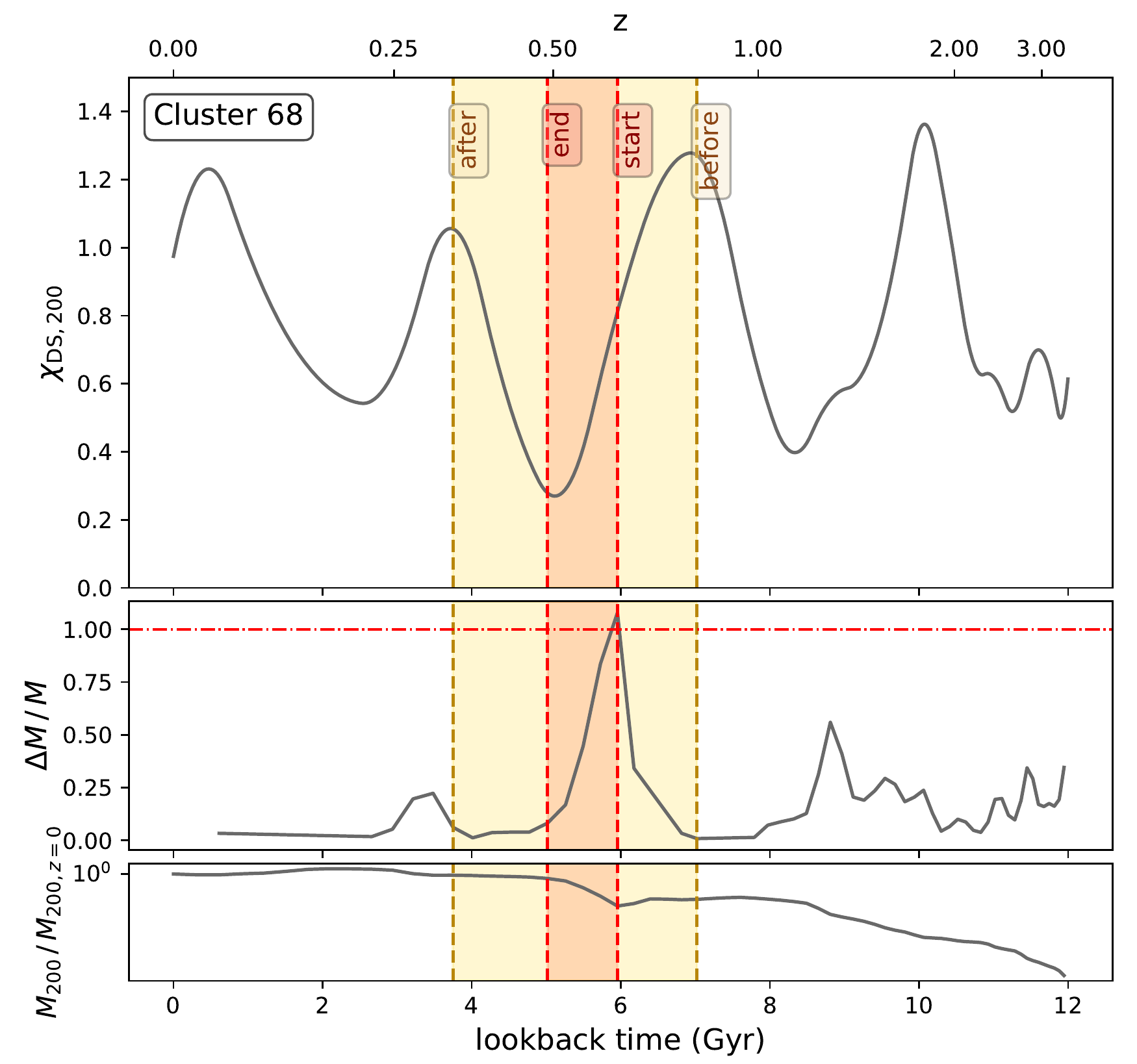}}
   {\includegraphics[width=8.5cm]{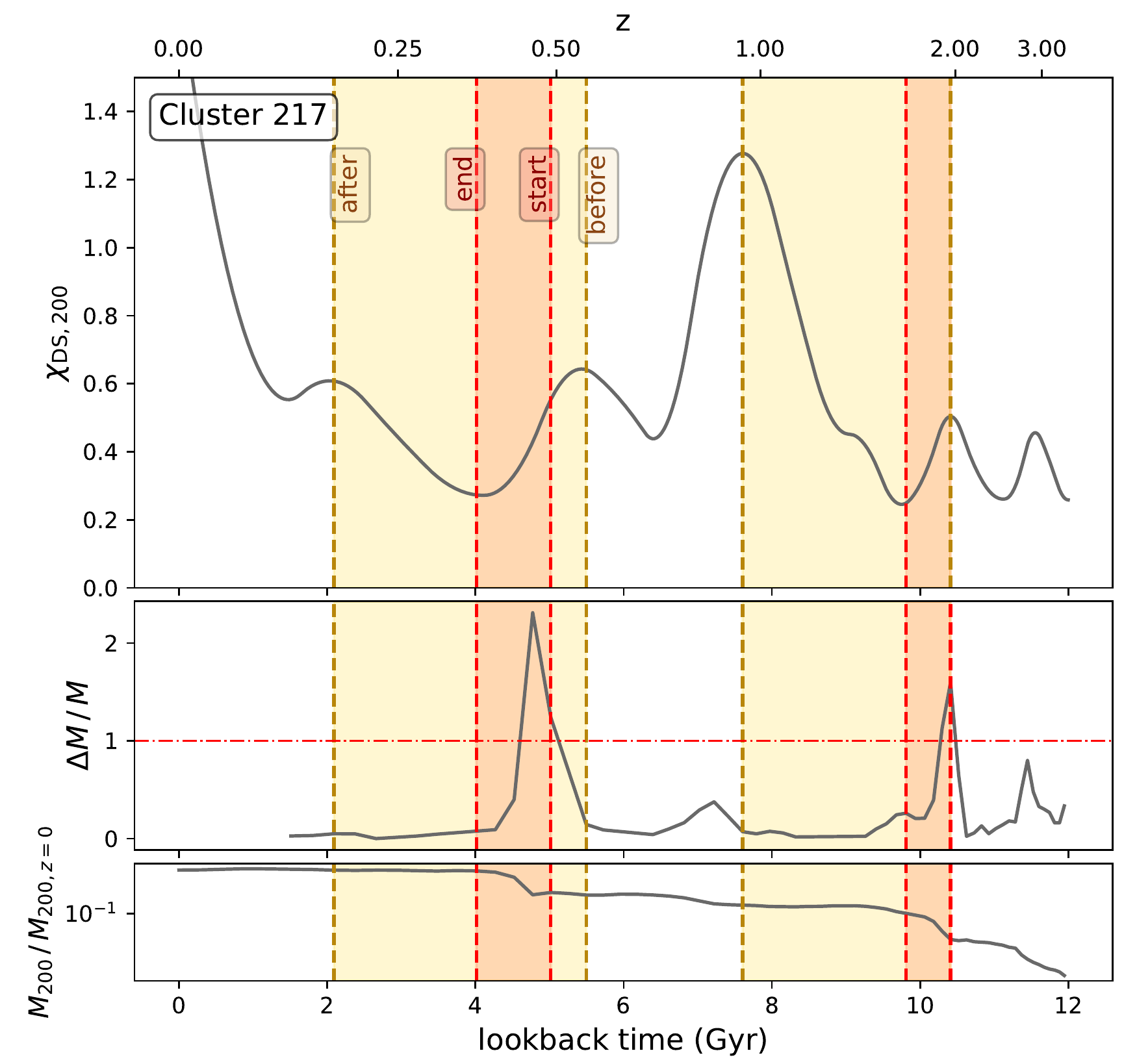}}
   \vspace*{-0.2cm}
   \caption{\textbf{Top panel}, relaxation parameter $\chi_\mathrm{DS,200}$ as a function of lookback time for two clusters selected as examples, 68 and 217. \textbf{Middle panel}, $\Delta M/M=(M_f-M_i)/M_i$, where the values in the x-axis correspond to the time $t_i$. \textbf{Bottom panel}, time evolution of $M_{200}$ in units of its value at $z=0$. We also provide the corresponding redshifts at the top of the plots. The horizontal line indicates the threshold used to identify mergers, i.e. $\Delta M/M = 1$. In all panels, the start and end point of the merger are indicated as red shaded regions that go (from right to left) from $z_\mathrm{start}$ to $z_\mathrm{end}$ (vertical red dashed lines). The yellow regions that overlap with them go from $z_\mathrm{before}$ to $z_\mathrm{after}$ (vertical brown dashed lines): we identify this particular region as `merger phase', i.e the time interval a cluster is disturbed by a merger.}
\label{fig:DeltaMDS}
\end{figure*}

For a cluster to be most relaxed, $\Delta_r$ and $f_s$ have to be minimised, and $\eta \rightarrow 1$, and so `dynamically relaxed' clusters have $\chi_{\mathrm{DS}} \gtrsim 1$. Moreover, a cluster with increasing $\chi_{\mathrm{DS}}$ is a relaxing cluster, whereas a sudden decrease in $\chi_{\mathrm{DS}}$ means that the cluster's equilibrium was disturbed. 

To gain a better understanding of this $\chi_\mathrm{DS}$ parameter, we have studied the influence of each of the individual parameters in the final value. In \citet{Neto07}, where these parameters were first introduced, the criterion of minimizing $\Delta_r$ is found to be the more restricting one; while in \citet{Power2012}, where both $\Delta_r$ and $\eta$ are studied, $\Delta_r$ is found to be the most robust measure of dynamical state in cosmological N-body simulations, showing a strong correlation with merger activity. This is also reflected in our \Eq{eq:chiDS}, where we have seen by plotting different $\chi_\mathrm{DS} (z)$ curves (not shown here though) that the parameter that carries the general shape, and therefore the one that contributes the most is the centre of mass offset $\Delta_r$. Regarding $f_s$, even though they follow different approaches, it depends on the position of haloes and subhaloes like $\Delta_r$ does, and so these two parameters show some correlation. On the other hand, the virial ratio $\eta$ makes use of kinematical information and thus probes the cluster in a different way. It can be proven to be responsible for the more dramatic changes in the $\chi_\mathrm{DS} (z)$ curve, rather than the general shape like the other two parameters. Given all this, we are confident that using the whole $\chi_\mathrm{DS}$ as in \Eq{eq:chiDS} gives robustness to our measurement of the relaxation parameter, while reflecting all the relevant information about the dynamical state of clusters; we will in turn use it now to quantify characteristic merger times.

\subsection{Characteristic merger times} \label{sec:characteristicmergertimes}
As mentioned before, mergers are events that extend over a certain period of time (according to our definition at most over half the cluster's dynamical time); neither the mass growth they produce nor the related effects they have on clusters are instantaneous. In this sub-section we now like to define four characteristic times of a merger:

\begin{itemize}
 \item[a)] $z_{\rm before}$ marking the time right before the merger takes place and when the cluster is still in equilibrium,
 \item[b)] $z_{\rm start}$ indicative of the onset of the merger (as defined via a pre-defined jump in its MAH),
 \item[c)] $z_{\rm end}$ marking the end of the actual merger, and
 \item[d)] $z_{\rm after}$ that characterizes the time when the cluster experiences a new relaxed phase. 
\end{itemize}

To find $z_{\rm start}$ we need nothing else but the MAH: we define it to be the redshift of the snapshot at time $t_i$ (see \Eq{eq:deltaM/M} above). For all other characteristic redshifts we additionally inspect the evolution of the dynamical state parameter $\chi_{\mathrm{DS}}$ as defined in the previous sub-section. Below we provide more (technical) details about the actual calculation of these four characteristic times.

In~\Fig{fig:DeltaMDS} we compare the dynamical state evolution of a cluster to its MAH for two representative regions. The top panel of this figure shows the evolution of $\chi_\mathrm{DS}(t)$\footnote{In this paper, $\chi_\mathrm{DS}(t)$ always refers to the parameter estimated inside $R_{200}$. Although the sub-index is indicated in \Fig{fig:DeltaMDS}, we drop it in the text for simplicity.}, whereas the two lower panels show the corresponding MAH as quantified by $(\Delta M/M)(t)$ (middle panel) and $M_{200}(t)/M_\mathrm{200,z=0}$ (bottom panel). For $\chi_\mathrm{DS}$, the curves shown are the result of taking a moving average of the values for every three snapshots and then interpolating, in order to filter the scatter. The curves show that, for every peak in $\Delta M/M$, there is a corresponding decrease in $\chi_\mathrm{DS}$ that then starts growing again, if there are no other similar events happening. As expected, there is a clear correlation between dynamical state and MAH, such that fast mass growths disturb clusters from their relaxed situations, but only for a short period before they return to a relaxed phase. Even though we only show two sample cases in \Fig{fig:DeltaMDS}, we confirm the same trends for all the clusters in our sample.

We further highlight in \Fig{fig:DeltaMDS} the aforementioned four characteristic times (with the corresponding redshift ranges as shaded regions, see figure caption for more details) whose definition in turn stems from the inspection of these curves for our clusters:

\begin{itemize}
    \item \bm{$z_\mathrm{start}$}: this time marks the beginning of the merger, which is the moment when the cluster starts growing in mass. This can be simply identified with the initial snapshot where the merger conditions are satisfied, obtained as described in \Sec{sec:definition}.
    
    \item \bm{$z_\mathrm{end}$}: this is the end point of the merger. To compute it, we could similarly use the final snapshot identified in the process of finding mergers, which, depending on the case, could be the snapshot that is strictly consecutive to the initial one, two snapshots after the initial one, or even longer if the merger is produced combining other mergers. However, this is only a measure of when the mass growth rate has fallen below the threshold set, which does not mean that the merger finished yet. For this reason, we use instead the dynamical state information and, since the relaxation of clusters decreases due to merger, we identify $z_\mathrm{end}$ with the first minimum in the $\chi_\mathrm{DS} (z)$ curve after $z_\mathrm{start}$. This is the moment the relaxation process restarts and the merger ends and its disturbing effects end, respectively.
    
    \item \bm{$z_\mathrm{before}$}: even before we reach $\Delta M/M=1$ the two merging objects start to influence each other. This is reflected in the decrease of $\chi_\mathrm{DS}$ prior to $z_{\rm start}$. Therefore,  $z_\mathrm{before}$ can be identified with the first peak in the $\chi_\mathrm{DS} (z)$ curve before the minimum which was identified as $z_\mathrm{end}$. 
    
    \item \bm{$z_\mathrm{after}$}: the last characteristic time is associated with the end of the whole merger phase, i.e. the time when the cluster returned to a relaxed position. As we have seen this is not the case for $z_{\rm end}$ which marked the start of the relaxation after the merger. We therefore define $z_\mathrm{after}$ to coincide with the first peak in $\chi_\mathrm{DS}(z)$ after the minimum $z_\mathrm{end}$. This is an indicator of when we can say that the cluster has `recovered' from the merger.
\end{itemize}

In summary, we have defined four characteristic times during a merger event, that can be computed for every merger using the MAH and dynamical state information of the cluster. Note that the start of a merger $z_{\rm start}$ is computed using only the MAH of the clusters, it does not include the mass ratio of the two most massive progenitors: this ratio enters later into our analysis (see \Sec{sec:mergers_massratio}, where we consider the masses of the main progenitors that belong to the merging clusters). The other three times are computed using dynamical state information $\chi_\mathrm{DS}$. Finally, since we are working with discrete snapshots in the simulations, their distributions will not be strictly continuous.

We further like to remark that our mass growth condition used in the definition of a merger is certainly a free parameter and we also investigated how our results are affected when lowering (or even increasing) it. Using different threshold values simply means shifting the dot-dashed line shown in the middle panel of \Fig{fig:DeltaMDS} up or down. Here one can clearly see that our choice rejects multiple minor mergers readily seen in the MAH as well as in the dynamical state parameter. Our methodology and merger time definitions will clearly hold for any such value. But for this particular study presented here we are primarily interested in how major mergers affect the internal properties of clusters and hence our choice of using mass growth by 100 per cent.

\subsection{Merger Sample}
Applying our MAH-based criterion to the full range of available cluster regions, we here specify the characteristics of the identified mergers a bit more. This includes quantifying the mass ratio of the two most massive progenitors, investigating the merger timescales, and defining a control sample that did not undergo a merger. We will also introduce a lower mass limit in order to only focus on mergers that are reasonably resolved in our simulations.

\subsubsection{Pre-selection of mergers} \label{sec:preselection}

We have described in detail our methodology to find and characterise mergers in our clusters sample. Before further analysing them we have to take into account the numerical resolution. First, we apply a cut in the identified value of $\Delta M/M$ for our mergers, and consider only instances where $\Delta M/M\leq 3$. This is to ensure that the mass growth seen is physical and not due to a misidentification by the halo finder (see \citealp{Behroozi15}). Then, to make sure that all of our mergers are well resolved, we introduce a lower limit, $N_\mathrm{cut}$, in the number of particles in the halo at $z_\mathrm{start}$, i.e. we discard all mergers where $N_\mathrm{part}(z_\mathrm{start})<N_\mathrm{cut}$. We will further investigate the effect of lowering $\Delta M/M$ from its usual value 1.

In \Tab{table:nmergers} we show the total number of mergers found for three different values of $N_\mathrm{cut}$; $1000$, $5000$ and $10000$, together with the values for no limitation, ``No $N_\mathrm{cut}$''\footnote{Note that the number of particles per object is limited to 20 as per mode of operation of \ahf.}. We can see how the number of found mergers decreases when increasing the cut in number of particles, but even for the most conservative choice of $N_{\rm cut}=10000$ (and the requirement to find a mass increase of a factor of 2) we still find more than 170 mergers across the whole sample. In previous convergence studies \citep{Trenti2010} it is shown that, for individual halo masses, $N_\mathrm{part} \gtrsim 5000$ is required to have an uncertainty smaller than 10$\%$ in the obtained value. Although some properties like the virial radius $R_\mathrm{vir}$ are very stable for low numbers of particles (see \citealp{Trenti2010}), other properties like halo shape \citep{Allgood06} or halo spin \citep{Benson2017a}, may require even more particles to have such a low uncertainty. As the prime objective of this work is to quantify the influence of mergers on internal properties of clusters, we prefer to apply the most conservative criterion and hence choose $N_\mathrm{part} \gtrsim 10000$ for the remainder of this work: comparing the columns for $N_\mathrm{cut}=5000$ and $10000$, we see that the differences between them are relatively small, and using $N_\mathrm{cut}=10000$ we are not losing that many mergers in our statistics. 

We have additionally included in \Tab{table:nmergers} possible variations of the other free parameter we have in our merger definition, the mass increase limit. 
Again, as we are primarily interested in major merger events, our usual choice is $\Delta M/M=1$, but to gain insight into how our statistics might increase when lowering this fractional mass increase we also provide the number of mergers for $\Delta M/M=0.75$ and $\Delta M/M=0.5$. We see that in the three cases the trends are the same, meaning that a lower mass growth criterion is only reflected in the total number of mergers, and not in their distribution regarding the number of particles in the halo.

\begin{table}
\centering
\caption{Total number of mergers depending on the different parameters chosen. Rows are different mass increase ratios, $\Delta M/M$, and columns show the threshold in the number of particles in the halo. Each threshold in $N_\mathrm{part}$ is also associated with a cut in redshift, where this number of particles is reached. The median redshift for each cut, computed including all the regions, is also indicated here.}
\begin{tabular}{l|c|c|c|c}
\hline
\multicolumn{1}{r|}{\textbf{}}    & \textbf{No} $\bm{\mathrm{N_{cut}}}$ & $\bm{\mathrm{N}_\mathrm{cut}=}$ \textbf{1000} & \textbf{5000} & \textbf{10000} \\ 
\textbf{median($z$):}     & 16.98      & 5.29                  & 3.93                  & 3.41                \\ \hline
\textbf{$\Delta M/M=1$}    & 575        & 339            & 225            & 178             \\ 
\textbf{$\Delta M/M=0.75$} & 875        & 533            & 368            & 308             \\ 
\textbf{$\Delta M/M=0.5$}  & 1440       & 885            & 666            & 560              \\ \hline
\end{tabular}
\label{table:nmergers}
\end{table}

In summary, with our selections, $\Delta M/M=1$ and $N_\mathrm{cut}=10000$ (which corresponds to approximately $M_{200} = 8\cdot 10^{12} \,\Msun$), we have a sample of 178 mergers, where we have also excluded situations where $\Delta M/M > 3$ to guarantee that the growth is physical. This way, even though we have been conservative in our choices to ensure the quality of the mergers, we end up with a sample that is still statistically significant.
 
\subsubsection{Mass ratio between merging haloes} \label{sec:mergers_massratio}
Using our MAH-based criterion to find mergers, we cannot yet say anything about the `type' of mergers, e.g. is it a `major' or rather a `minor' merger. Therefore, and following other works  \citep[e.g.][]{Planelles2009,ChenImprints2019}, a quantity that is very interesting to characterise mergers is the mass ratio between the merging objects, which allows for the classification into minor ($0.1 \leq M_2/M_1 < 0.33$) and major mergers ($M_2/M_1 \geq 0.33$), where $M_2$ and $M_1$ are simply the masses of the two main progenitors, chosen so that this ratio is always $\leq 1$.
To obtain this ratio we use the trees created with \textsc{MergerTree} to find a list of all the progenitors of the central halo at the initial snapshot of the merger. To find the progenitor that is merging with the main progenitor, we select the one that maximizes the function $M_\mathrm{prog,i}/M_\mathrm{prog,1}$, where $M_\mathrm{prog,1}$ is the mass of the main progenitor and $M_\mathrm{prog,i}$ the mass of each of the other progenitors. 

\begin{figure}
   \hspace*{-0.1cm}\includegraphics[width=8.5cm]{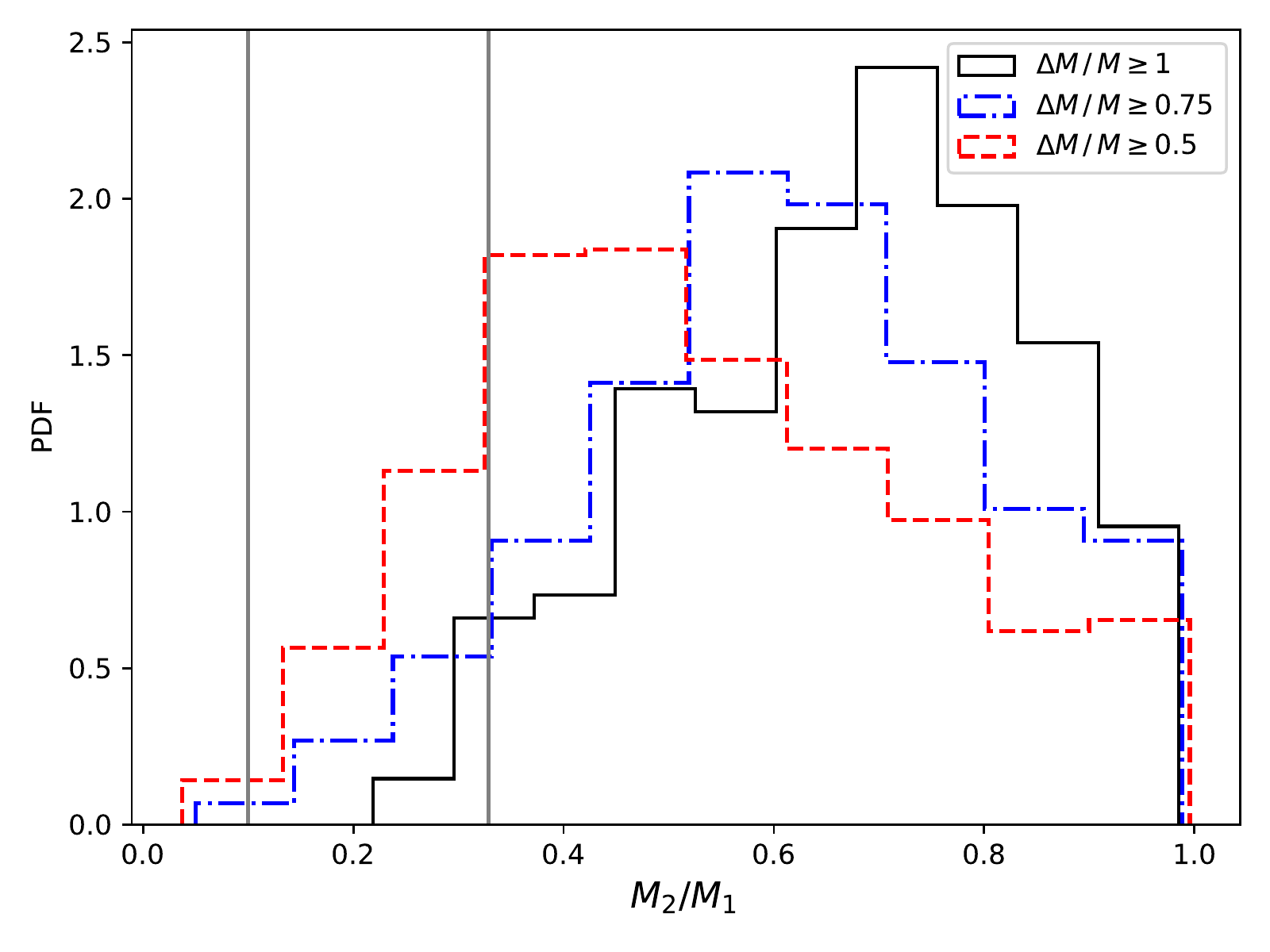}
   \caption{Distributions of mass ratio between the two merging haloes, $M_2/M_1$, computed as described in \Sec{sec:mergers_massratio}. The plot compares the distributions for the three different mass increase ratios indicated, all with the limit $N_\mathrm{cut}=10\,000$ in the number of particles in the halo. The vertical lines designate the limits for the definitions of minor and major mergers (0.10 and 0.33).}
\label{fig:MassRatio21}
\end{figure}

\Fig{fig:MassRatio21} shows the distribution of this value for all the mergers in our sample, for the three different $\Delta M/M$ thresholds. For $\Delta M/M=1$, it can be seen that almost all the mergers are major mergers ($M_2/M_1 > 0.33$), with a peak in the distribution at around $M_2/M_1=0.7$. This is reasonable, since the main branch halo is increasing its mass by 100 per cent, and thus we expect it to be merging with a very massive halo (and the remaining mass coming from minor mergers and accretion). Looking at the other values of $\Delta M/M$, we see that the distributions are shifted towards lower $M_2/M_1$ ratios, meaning that the objects merging with the main branch are now smaller -- as expected. However, the differences between the three distributions are quite small, and even for the $\Delta M/M=0.5$, where we have 560 mergers (see \Tab{table:nmergers}), we can see that most of them are classified as major mergers. In spite of this, for further calculations we are still going to use the $\Delta M/M=1$ sample, since this is where we get the highest number of major mergers, which is the scenario we are most interested in for this particular study.

We like to restate that our `merger' definition only uses the MAH of the clusters (see \Sec{sec:definition}), and hence the situations we find and classify as `mergers' are in fact just very fast and significant mass growths. In this subsection, by comparing the mass of the main and the (second) most massive progenitors of the main halo, we have verified that these mass growths can in fact be associated with (major) mergers, as demonstrated in \Fig{fig:MassRatio21}. For the cases where the ratio $M_2/M_1$ is lower, we have to take into consideration that there are other (less frequent) situations that can lead to a significant mass growth with a low $M_2/M_1$ ratio associated. These include multiple-object mergers or simple accretion of material, which we are not considering here. 

\subsubsection{Merger timescales} \label{sec:mergertimes}

Another interesting quantity to calculate is the timescale of the merger. Using our characteristic merger times defined above in \Sec{sec:characteristicmergertimes}, we have two options, i.e.

\begin{itemize}
    \item $z_{\rm end}-z_{\rm start}$: the length of the actual merger event (referred to as `time of merger'), and
    \item $z_{\rm after}-z_{\rm before}$: the `merger phase', i.e. the time it takes to go from the unperturbed pre-merger state to the new post-merger state.
\end{itemize}

\begin{figure}
   \hspace*{-0.1cm}\includegraphics[width=8.5cm]{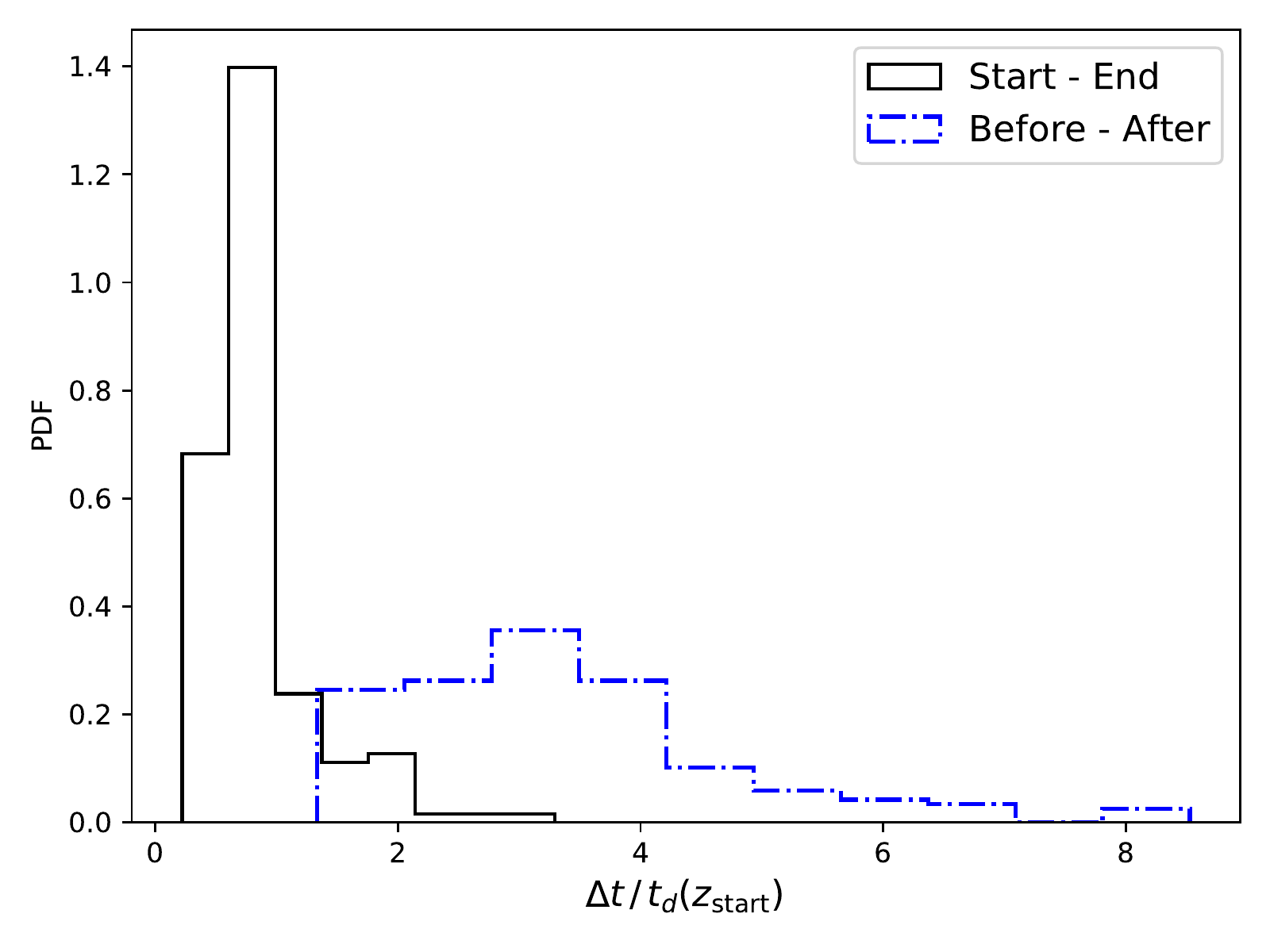}
   \caption{Distributions of $\Delta t = t_\mathrm{snap,i}-t_\mathrm{snap,f}$ in units of the dynamical time, $t_d$, at $z_\mathrm{start}$. The two different timescales go from $z_\mathrm{start}$ to $z_\mathrm{end}$ (solid black) and from $z_\mathrm{before}$ to $z_\mathrm{after}$ (dash-dotted blue).}
\label{fig:Deltat}
\end{figure}

To quantify the length of these periods, we convert the difference in redshift to $\Delta t$ which will further be normalized by the dynamical time $t_d$ at the larger redshift \footnote{To give the reader an approximate idea of the actual values of these timescales, the dynamical time, $t_d$, of clusters (which depends only on redshift, see \Sec{sec:definition}) ranges from $\sim$1.4 Gyr at $z=0$ to $\sim$0.5 Gyr at $z=2$.}. \Fig{fig:Deltat} shows the distribution of $\Delta t/t_d$ for the two different time spans. The plot is based upon our usual sample (i.e. $\Delta M/M=1$ and $N_{\rm cut}=10000$), but only includes the results for all the mergers for which the four characteristic times could be computed: mergers that happened very recently may have not had enough time to reach either the end of the merger or to relax again. This happens for 14 mergers leaving us hence with 164 for that plot (our `reduced' merger sample). In the figure it can be seen that the distribution of the time of the merger peaks at $\Delta t \sim 0.7 t_d$, which could be expected given our merger definition, that uses $0.5 t_d$ to search for mass jumps. 
For the merger phase, the distribution peaks at substantially higher times ($\sim 3 t_d$) and is also much wider, with values up to $8 t_d$. This means that, while mergers are rather rapid, returning to equilibrium takes clusters quite some time. 

We have further correlated the time of merger with the merger phase (though not explicitly shown here) where we find that the Spearman correlation coefficient between these two quantities is $r_s \simeq 0.4$, which shows that, in general, clusters that undergo a long lasting merger, also take more time to relax afterwards, as expected.

\subsection{Control Sample}  \label{sec:controlsample}

Now that we have defined and characterised mergers, we have a very well-defined merger sample. But before proceeding to study the effect mergers have on different cluster properties, we require a `control sample'. Simply comparing cluster properties before and after the merger is not sufficient for drawing any conclusions about the effect of mergers on clusters. We need to compare our results against a sample of clusters that did not undergo (comparable) merger. For this reason, we need to define a control sample to be compared against.

To create the control sample, we first obtain the values of $z_\mathrm{start}$ and $M_\mathrm{start} = M(z_\mathrm{start})$ for each merger in the merger sample. Then we check each of the other 323 clusters in \textsc{The Three Hundred} sample, and select all those that, at $z_* = z_\mathrm{start}$, have a mass within 1$\%$ of $M_\mathrm{start}$, i.e., those which satisfy $\vert M(z_*) - M_\mathrm{start}\vert < 0.01 M_\mathrm{start}$. This way we obtain a list of objects that has a comparable mass distribution at $z_\mathrm{start}$ as the merger sample, and also the same distribution for $z_\mathrm{start}$.

We further need to make sure that -- within the same time span as characterized by the `merger time' -- those objects do \textit{not} undergo a merger. In order to achieve this, we first need to set the corresponding characteristic times. Since the values of $z_\mathrm{start}$ are already defined, we need to define the other three characteristic redshifts. Focusing first on $z_\mathrm{end}$, our aim is to mimic the distribution of the length of the time of the merger itself. For that, we obtain the distribution $z_\mathrm{end}-z_\mathrm{start}$ for the merger sample and resample from there. This is, for each item in the control sample we pick a random value from this distribution for the difference between these two times. We then compute the value of the control $z_\mathrm{end}$ based on $z_\mathrm{start}$. We can just repeat this process for $z_\mathrm{before}$. For $z_\mathrm{after}$, since we want to also mimic the distribution of the length of the whole merger phase, we resample from the distribution of $z_\mathrm{after}-z_\mathrm{before}$, and derive the value of the control $z_\mathrm{after}$. As a result, we obtain a control sample consisting of a list of clusters with four characteristic times corresponding to the values found for the merger sample.

We finally need to ensure that our objects selected this way do not undergo a merger. For this purpose, we check the value of $\Delta M/M$ for every instance in the control sample. We obtain this value by computing the value for each snapshot from $z_\mathrm{before}$ to $z_\mathrm{after}$ and then selecting the maximum. The mergers in our merger sample were identified following the criterion $\Delta M / M \geq 1$ and so, to set an even more restrictive criterion, we verify that every instance in our control sample satisfies $\Delta M/M<0.8$. Besides, we also check that they satisfy the resolution condition $N_\mathrm{part}(z_\mathrm{start})>N_\mathrm{cut}$ (see \Sec{sec:preselection}). We discard every instance where any of these two conditions is not satisfied, obtaining our final control sample.

It is important to note that this method has a random component due to the random resampling of the redshifts, and thus a different control sample can be obtained in each execution. For consistency, we are going to work with a fixed random seed, so that our sample is always the same for all the calculations (we have checked different samples and got essentially the same results for all). With this certain seed, we have created the control sample described, and obtained a list of 250 items, which mimic the mergers in our merger sample, but where no significant mass growth is happening. 

\begin{figure}
   \hspace*{-0.1cm}\includegraphics[width=8.5cm]{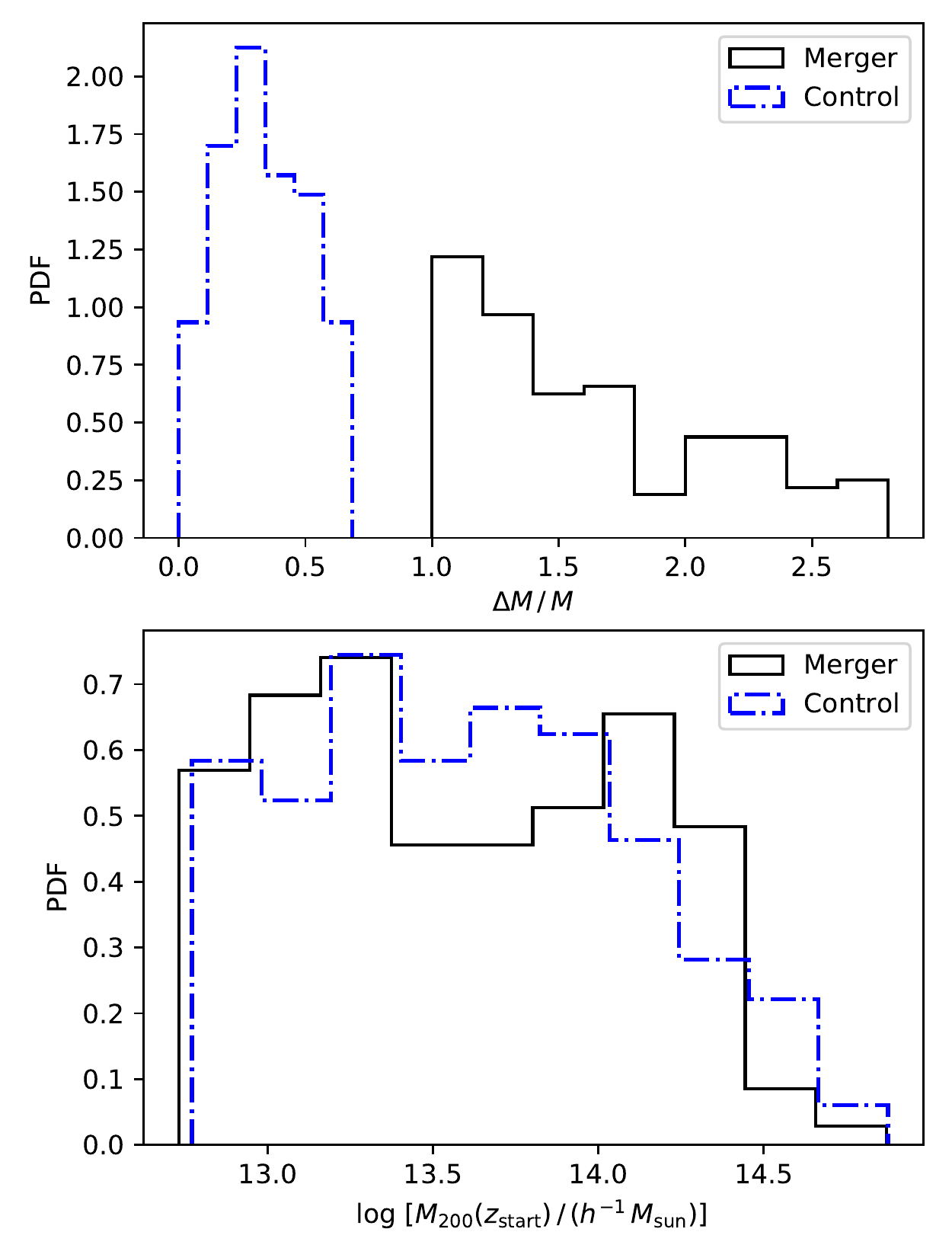}
   \caption{\textbf{Top,} distributions of the value of $\Delta M/M$ for the merger sample (solid black) and the control sample (dash-dotted blue) described in \Sec{sec:controlsample}. \textbf{Bottom,} distributions of the mass of the cluster at $z_\mathrm{start}$ for the same two samples.}
\label{fig:ControlsMass}
\end{figure}

As a confirmation of this, the top panel of \Fig{fig:ControlsMass} shows the distribution of the value of $\Delta M/M$ for the mergers in the merger sample, together with the same values obtained for the control sample. As we have imposed, the values in the control sample range from 0 to 0.8, peaking around the centre of this range, whereas the merger sample peaks at 1 (which is a lower limit due to our choice), but shows a wider and decreasing distribution until the cut made at 3. The bottom panel of \Fig{fig:ControlsMass} shows the distribution of the mass of the cluster at $z_\mathrm{start}$, again comparing the merger sample with our control sample. In this case we can see that the distributions are similar, the range of masses being the same for both. This is further confirmed by conducting a two-sample Kolmogórov-Smirnov (K-S) test, which yields a p-value of 0.40 that implies that the two distributions could have been drawn from the same parent distribution. This way we ensure that there will be no mass bias in future comparisons between the merger and the control sample. Note that in these plots we are working again with the `reduced' samples, i.e. excluding the situations where the merger is not over yet (and thus $z_\mathrm{end}$ or $z_\mathrm{after}$ are not found). This leaves us with 164 mergers in the merger sample and 236 items in the control sample. 

\section{Effect of mergers on BCGs} \label{sec:results} 

We will now focus on the brightest cluster galaxies (BCGs) -- and particularly on their stellar component -- studying how mergers are affecting them. Based on the previously described merger and control samples, we investigate properties like colour and luminosity and assess the impact of (major) galaxy cluster mergers on them. This is relevant for understanding the evolution of these giant galaxies in the context of the hierarchichal formation scenario.

\subsection{Stellar mass in BCGs} \label{sec:results_stellarmass}

A simple way to analyse the stellar component is by measuring the mass in stars in the central galaxy. Since the edge of a BCG is not clearly defined, we determine it by selecting all the stars inside a certain radius, using the halo centre as the origin. We call this parameter the radius of the central aperture, $R_{ca}$, and we employ three different values for it: 30, 50 and 70 kpc, so that we can study the dependence of the results on the region considered. This way of defining BCGs, based on a fixed physical (not comoving) radius, has often been adopted in simulations. For instance, \citet{McCarthy2010} used a radius of 30$\hkpc$, whereas in \citet{Ragone-Figueroa2018} three different radii were used too (30 kpc, 50 kpc and $0.1R_{500}$). \citet{Kravtsov18} advocate the use of stellar masses defined this way for comparison with observations, and they provide values for nine BCGs using several different radii. Besides, \citet{Stott2010} showed that, using a 50 kpc aperture radius, BCG luminosities can be recovered with less than 5 per cent difference from those obtained in some observational analyses. Hence, using $R_{ca}$ allows us to easily count and identify in the simulation the stellar particles that are in this central region at every stage of the merger. 
We like to remark that we are only going to use this definition at the characteristic merger times previously described. This way, although because of the mergers the mass can be ill-defined by this approach in-between these times, the halo centre is always stable at these times, and thus the BCG is well identified.

\begin{figure}
  \hspace*{-0.1cm}\includegraphics[width=8.5cm]{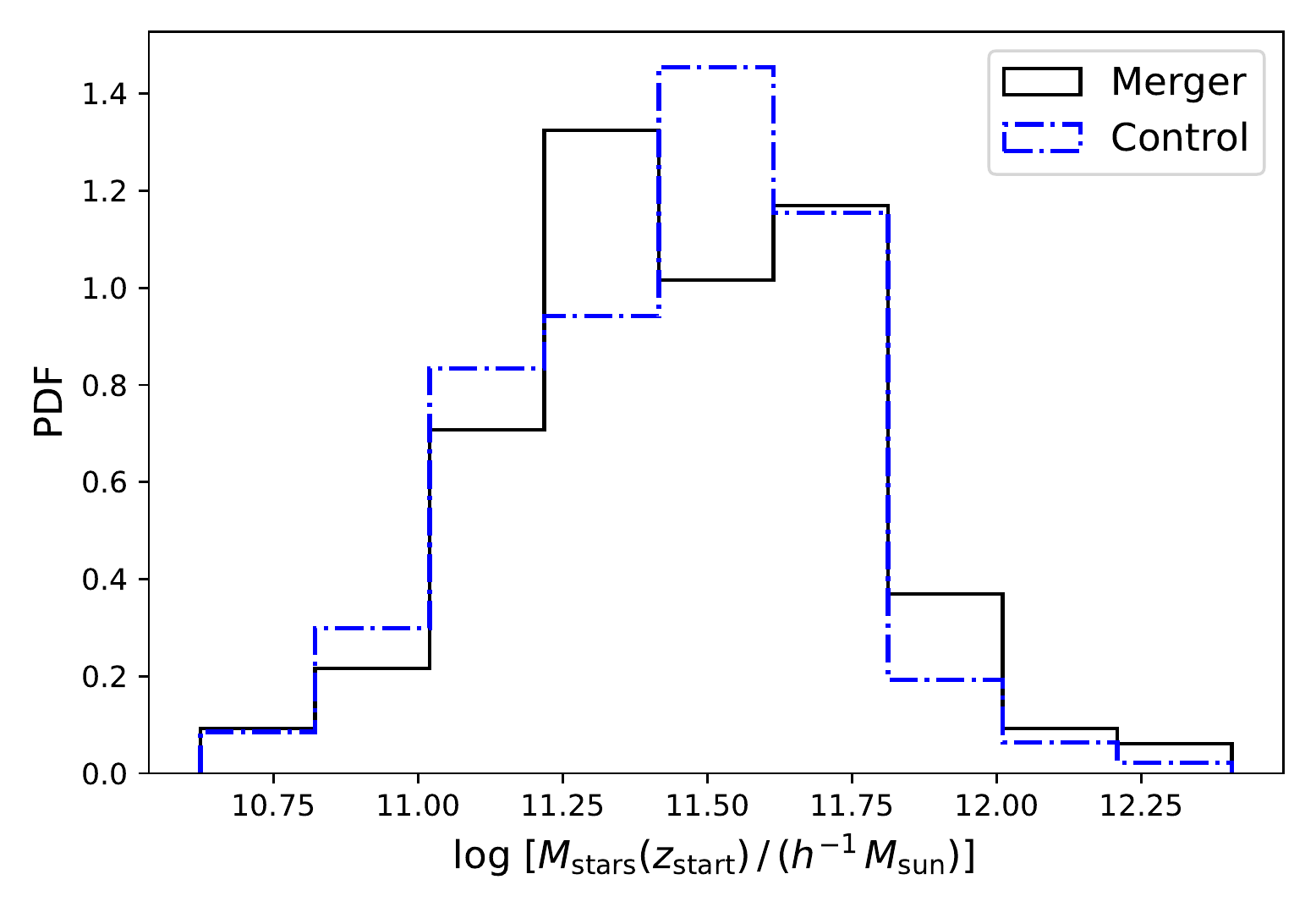}
  \caption{Distribution of the stellar mass of the BCG at $z_\mathrm{start}$ for $R_{ca}=50$ kpc. Values for the merger sample (solid black) and the control sample (dash-dotted blue).}
\label{fig:BCGmass}
\end{figure}

As a way to check the obtained values and compare with the halo mass (shown in \Fig{fig:ControlsMass}), we have plotted in \Fig{fig:BCGmass} the distribution of the stellar mass of the BCGs at $z_\mathrm{start}$, only focusing on $R_{ca} = 50$ kpc for simplicity. In this figure, similarly to the lower panel in \Fig{fig:ControlsMass}, we compare the distributions for the merger and the control sample. We confirm again that our control sample is not biased towards more (or less) massive BCGs.

\begin{figure}
   \hspace*{-0.1cm}\includegraphics[width=8.5cm]{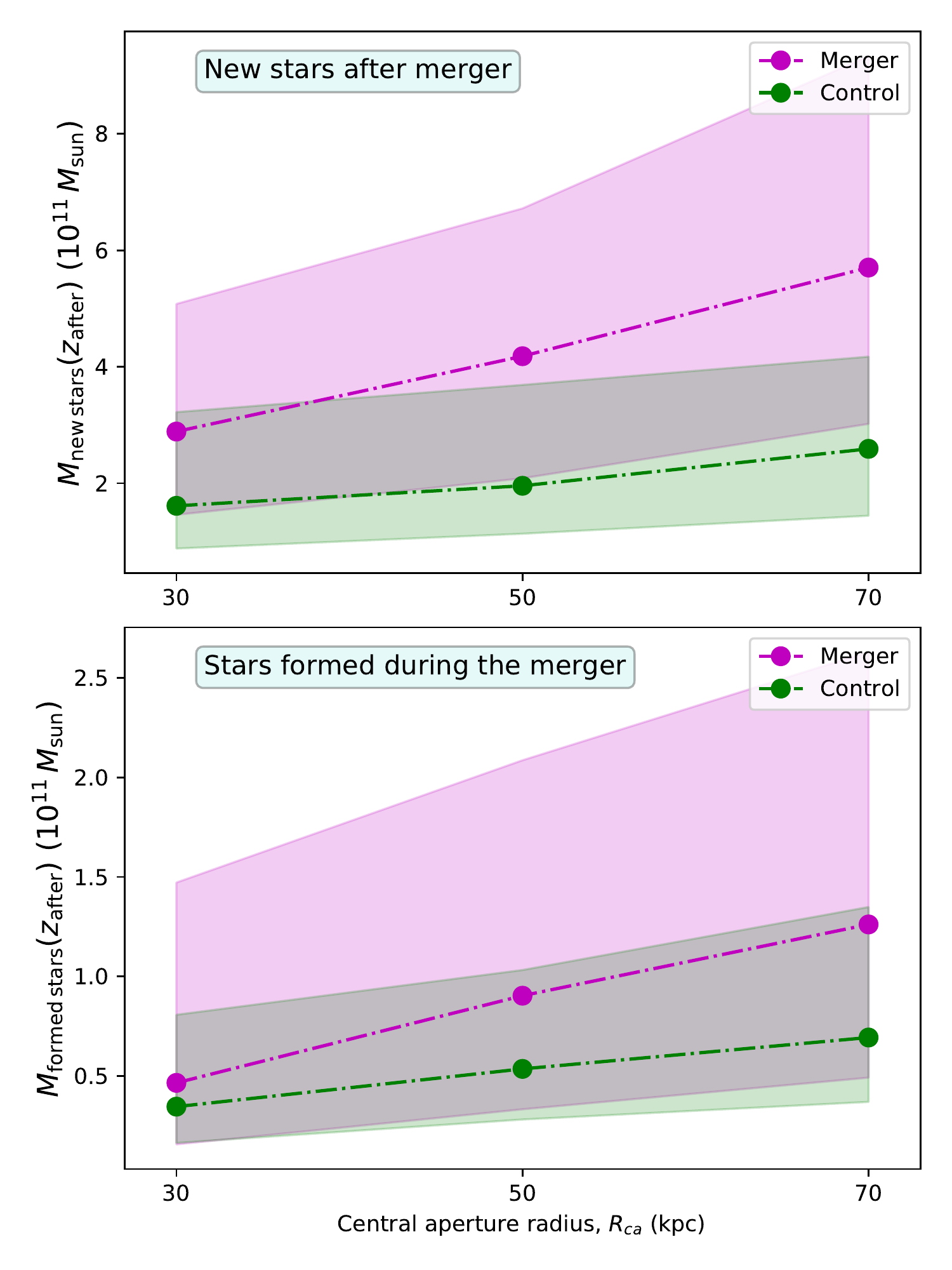}
   \caption{\textbf{Top,} median `new stellar mass' in the BCG of the clusters, defined by the physical aperture radius $R_{ca}$. This mass is defined as the sum of the mass of all the stellar particles identified at $z_\mathrm{after}$ that were not identified at $z_\mathrm{before}$. In magenta the median for all the mergers in the merger sample, and in green for the control sample. Shaded regions indicate the 25 and 75 percentiles. \textbf{Bottom,} median stellar mass that was formed during the merger. Computed as the sum of the new stellar particles whose age is smaller than the length of the merger phase.}
\label{fig:new_stars}
\end{figure}

To study the effect mergers have on BCGs, we analyse the stellar components before and after the merger (see merger times as defined in \Sec{sec:mergertimes}), comparing them to the control sample where no particularly fast mass growth is taking place.
In \Fig{fig:new_stars} we quantify the mass in new and newly formed stars as a function of the applied $R_{ca}$ for both the merger and control sample. The upper panel shows the values for the mass of new stars (i.e. stars that did not belong to the BCG prior to the merger) that are identified in the central region (BCG) of each cluster after a merger. To obtain this quantity we identify all the stellar particles whose ID is identified at $z_\mathrm{after}$ in the corresponding central region, but not at $z_\mathrm{before}$, and then we sum all their masses. Note that the number of `new stellar particles' is not necessarily the difference between the number of stars before and after the merger. We explicitly count new stars that previously were not inside $R_{ca}$, not accounting for possible stellar mass loss (which we find to be negligible, see discussion below).
We compute this mass for all the clusters in each sample and then plot the median values (green and magenta dots) and the 25 and 75 percentiles (shaded regions). We find that, as expected, the stellar mass of BCGs grows more in mergers than in the control sample, especially when considering the outer regions up to $R_{ca}=70$ kpc.

Here we also considered the stellar mass that is `lost' during the merger, in the form of stellar particles that leave the central region. However, we find this quantity to be nearly independent of the radius $R_{ca}$ considered. The median value of this `lost mass' is $\sim 0.8 \cdot 10^{11} \Msun$ for the merger sample and $\sim 0.6 \cdot 10^{11} \Msun$ for the control sample. Comparing these values to those in the upper plot of \Fig{fig:new_stars} we can see that the mass of new stars is much greater, confirming that BCGs are growing significantly during this time. 

The lower panel of \Fig{fig:new_stars} shows the median stellar mass \textit{formed} during the merger in each BCG. This was computed by comparing the age of all the new stellar particles with the length of the merger phase in Gyr. The plot has the same features as the upper panel plot, i.e. we see more star formation in the merger sample, specially for $R_{ca}=50$ and $70$ kpc, where the difference is around 70 per cent with the control sample. Comparing the quantities in both plots, we note newly formed stars only contribute in $\sim 20$ per cent to the total stellar mass growth.
This leads to the conclusion that the main reason for BCGs to grow during a merger is the accretion of existing stars, although we also appreciate that actual star formation takes place during the merger.

\begin{figure*}
   {\includegraphics[width=17.5cm]{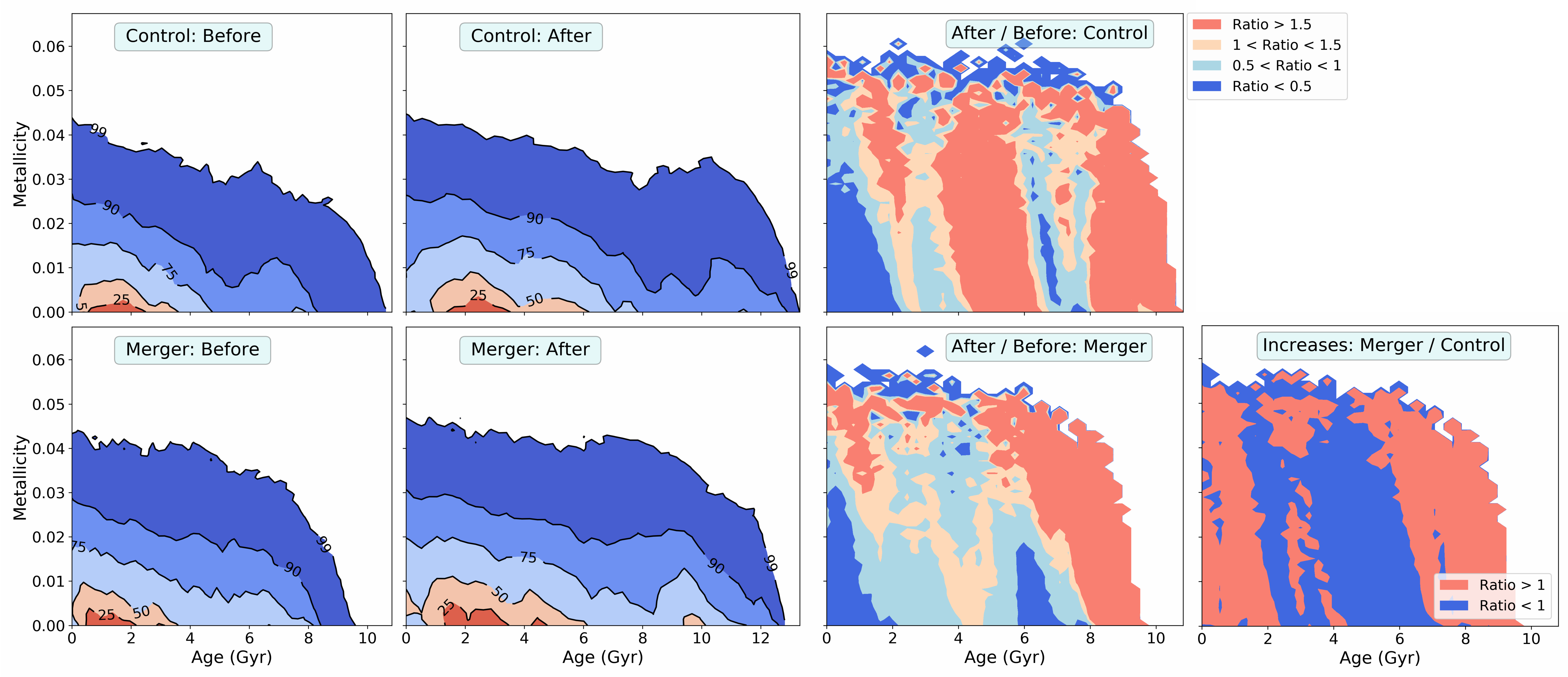}}
   \caption{2D histogram of the (mass-weighted) age and metallicity of all the BCG stellar particles for the indicated sample and time. For the first two columns on the left: first column for the time before the merger, second column after. Top row for the control sample and bottom row merger sample. The two panels on the third column show the ratios between the two times (after over before) for the control (top) and merger sample (bottom). The last plot, on the bottom right corner, shows the ratio between the two plots on the third column, i.e., the ratio after/before for the merger sample over the same ratio for the control sample.}
\label{fig:2dhists}
\end{figure*}

Finding both that stars are being lost and formed during the merger, it raises the question if the two partaking BCGs actually merge to form a new BCG. To shed light into this, we have compared the two BCG masses before the merger and the resulting BCG mass after the merger. We found that the sum of the pre-merger BCG masses in general only accounts for about 60 per cent of the new BCG mass. This is in line with our previous finding that the new BCG contains in fact new stars. We further confirm that in most of the cases one of the two merging BCGs contributes nearly all of its stars with the other supplying only a small fraction, whereas in some cases no merger between the BCGs occurs at all. In the end we deduce that the situation is not as simple as `two BCGs merge to form a new one'. There are stars accreted from the surroundings, new stars are being formed, and this happens primarily in one of the partaking BCGs.

An important concern emerging at this point relates the exact particulars of the BH treatment, which has been shown to have relevant consequences not only on BH properties themselves, but also on other components of the cosmological simulation. For instance, \citet{Bahe2021} study the influence of BH repositioning on baryonic components of galaxies. A different work by \citet{Ragone-Figueroa2018} found that an improved `BH pinning' can result in smaller BCG masses, eventually leading to better agreement with observations. \citet{Bassini2020} emphasise that BHs being moved from the centre is a relevant problem in cosmological simulations, in particular for galaxy clusters, where the absence of AGN feedback -- due to exactly such `BH displacement' -- can produce non-physically high star formation rates. As merger events can be drivers for such displacements, we have therefore carefully investigated any possible correlation between them and the star formation observed during the merger. We find that during the merger the overall BH mass inside the considered BCG aperture increases, but not necessarily right at the centre. There is a median increase in the BH distance to the centre of $2.0, 4.5$ and $7.8$ kpc for the $30, 50$ and $70$ kpc apertures, respectively \footnote{For more detail, the first and third quartiles of these distributions are, respectively: -1.8 and 7.1, -2.1 and 12.3, -3.3 and 15.7 kpc. The negative values here indicate that the BH is becoming closer to the centre.}. Such an increase is not found for the control sample though. We further cannot confirm any correlation of this increase with the observed rise in newly formed stars during the merger. But we also need to remark that the overall BH mass increase mentioned here not only refers to a mass growth of an individual BH, it could also mean that additional BHs have been captured by the BCG. They either directly come from the merger component(s) or constitute BHs lost by satellite galaxies during their passages near the BCG. In any case, there are hardly ever more than 2-3 BHs inside a BCG -- irrespective of the applied aperture -- and their velocity distribution follows the same as that of the stars making up the BCG. Considering the numerical challenges in holding the BHs at the centre of their galaxies during a merging process, it is not surprising to find, in some cases, that several BHs were not merged into a single one.

\subsection{Stellar age and metallicity}

As we eventually aim at studying the effect of mergers on luminosities and colours, we will first investigate here those values that are being used by \textsc{stardust} to calculate these properties, i.e. stellar mass, age, and metallicity. By conducting this study we make sure we understand the properties of our BCG samples as given by the simulation before using them to derive more observable properties in post-processing. We focus again on stellar particles as found within a fixed aperture around the halo centre. 

Our results are summarized in \Fig{fig:2dhists} where we show in the first $2\times2$ plots the mass-weighted distributions of metallicity and age for the control  (top row) and merger (bottom row) sample before (first column) and after (second column) merger. We chose to show 2D histograms as the pair age-metallicity is relevant for the selection of the spectra in \textsc{stardust} (and the stellar mass is also used as a weight in that code). For simplicity, we only show here the values for the BCGs obtained using $R_{ca}=50$ kpc, but similar results are obtained for the other two apertures. By comparing the first and second columns, the aging of the stars can be seen in both samples. This corresponds to the time elapsed during the merger phase, from `before' to `after' the merger: looking at \Fig{fig:Deltat} in \Sec{sec:characteristicmergertimes}, we can see that the peak of this distribution is at $3\times t_d$. For $t_d \sim 1$ Gyr, the aging is around 3 Gyr, in agreement with what is shown in the first two columns in \Fig{fig:2dhists}.

The temporal evolution of the 2D histograms has been quantified in the third column, which shows the ratios between the distributions after and before the merger for the control (top) and the merger (bottom) samples. The blue regions, with `Ratio < 1', indicate a decrease of stars for these pairs of age and metallicity, while we find an increase in the red regions. 
Note that the ratios are between the normalised histograms, rather than the absolute number of star particles. This way, a ratio greater than 1 means that the percentage of stars in a sample with that particular age-metallicity is increasing. If the histograms were not normalised, the ratio would be > 1 in most of the regions, since, in general, the number of stars in the BCG is increasing during the merger, as we saw in the previous subsection.

Looking in detail at the third column in \Fig{fig:2dhists}, the main finding is indeed the aging of the stars. For both samples, the percentage of older stars is higher after the merger, specially above 8 Gyr. Focusing now on the younger stars, with ages below 2 Gyr, we see that the situation is very similar for both samples, with the percentage of stars being higher before the merger than after (blue region in the histograms). There is only a slight difference in the region with ratio < 0.5 for ages below 2 Gyr, with this region being smaller for the merger sample (bottom row). This can be interpreted as a hint of what we saw in the previous subsection, that BCGs in the merger sample are actually forming more new stars, and thus the difference for these young ages with the control sample. To further observe this effect, the plot in the bottom right corner of \Fig{fig:2dhists} shows the ratio between the two plots of the third column. That is, it shows the ratio between the change in the stellar distribution for the stellar particles in the merger sample over the same change for the control sample. We observe that for all metallicities, the increase in stars younger than 2 Gyr is greater for the merger sample, which also validates our previous findings.

To summarize this subsection, we have studied the mass-weighted age-metallicity distributions of all the BCG stellar particles, comparing the merger and the control samples before and after the merger. We have only found some small differences between them that nevertheless can have an impact on observable properties like colour and luminosity, something we will analyse in the following sub-subsection.

\subsection{Colour and luminosity}

\begin{figure}
   \hspace*{-0.1cm}\includegraphics[width=8.5cm]{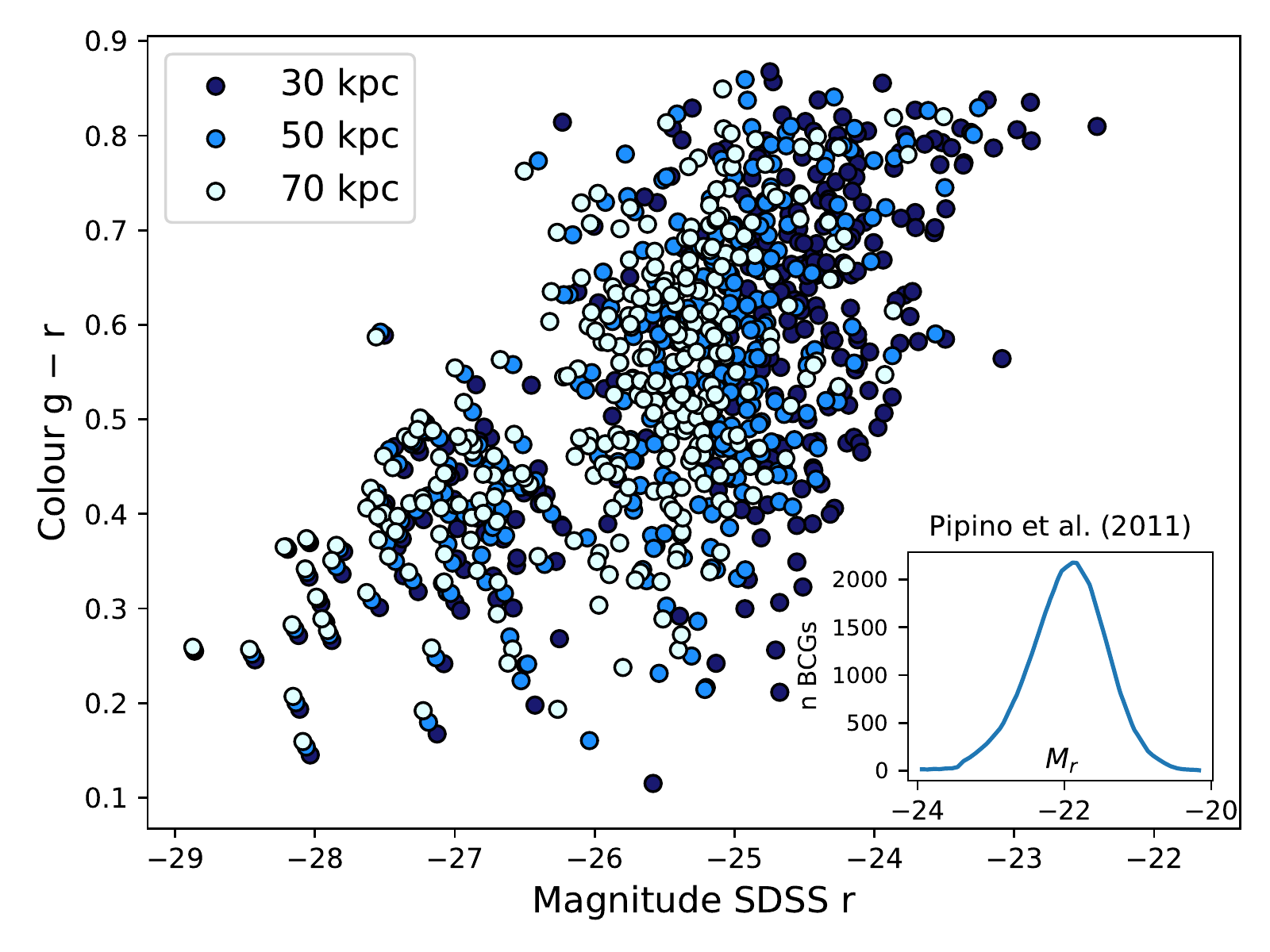}
   \caption{Colour-magnitude diagram for the BCG of the 324 clusters in \textsc{The Three Hundred} sample at $z=0$. The results for the three different values for $R_{ca}$ are shown: 30 (dark blue), 50 (light blue) and 70 kpc (white). The inset shows the distribution of BCG magnitudes for the observational SDSS sample from \citet{Pipino2011}.}
\label{fig:cmdz0}
\end{figure}

 \begin{figure*}
  \hspace*{-0.1cm}
  \includegraphics[width=15cm]{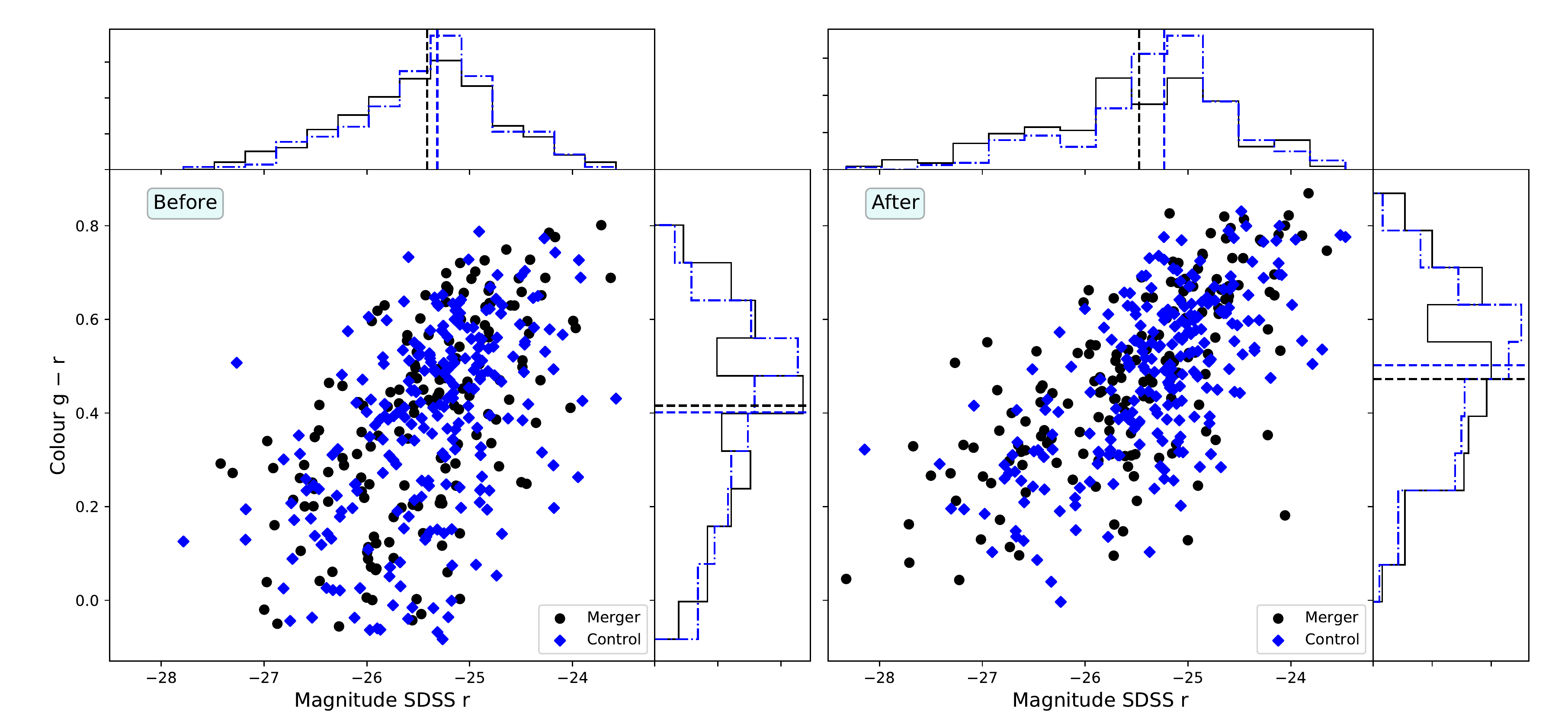}
  \caption{Comparison of colour-magnitude relations for the merger (black dots) and the control sample (blue diamonds). The values are for the BCG at $z_\mathrm{before}$ on the left and at $z_\mathrm{after}$ on the right. The distributions of the $g-r$ colour and the SDSS$-r$ magnitude are also shown in the plot for each sample (black solid lines for merger and dash-dotted blue for control) and time, together with their median values.}
\label{fig:cmds}
\end{figure*}

In the two previous subsections we have seen that the stellar component of BCGs is growing significantly more in the merger sample than in the control sample. Moreover, we have found that mergers produce a burst in star formation, which we have confirmed by looking at the distribution of all the stellar particles in age and metallicity. Although we have seen that these bursts are, in general, weak (the mass formed is a small fraction of the final stellar mass, leaving a small imprint on the age-metallicity distributions), previous studies \citep{Perez-Gonzalez2003} have shown that, even for bursts as weak as 1 per cent, the young population has a major influence on the optical colours of galaxies. 

For this reason, in this subsection we are going to analyse the impact mergers have on observable properties of the BCG. As new stars are formed during the merger, we might find an effect on luminosity and -- in particular -- colour, since young stars are hot and blue, as opposed to the red colours of the older stars. In order to check this, we can use the stellar code \textsc{stardust} again \citep[][see also \Sec{sec:data}]{Devriendt99}, but this time \textit{only} applying it to the stars found in the BCG. \textsc{stardust} uses the given age and metallicity of each stellar particle and obtains from a catalogue the full spectrum of a star with those properties. Then, each spectrum is weighted by the mass of each particle, and the sum of all of them yields the galaxy SED. This way, we can compute the luminosities and magnitudes for the BCG stars, considering all the identified stellar particles within each $R_{ca}$ region at each snapshot, with their properties we already analysed. In particular, we have used this code to obtain the luminosities in the SDSS bands, and from the magnitudes we compute the widely used $g-r$ colour index.

\subsubsection{Colour-Magnitude Diagrams}
Before analysing changes in luminosity and/or colour due to mergers, we first present our absolute values. In \Fig{fig:cmdz0} we display a $g-r$ colour-magnitude (at SDSS-$r$ band) diagram for all the BCGs in \textsc{The Three Hundred} sample at $z=0$, computed for the three values adopted for $R_{ca}$. We see that the values for 70 kpc (in white) are in general brighter than those for $R_{ca} = 50$ (light blue) and 30 kpc (dark blue), which is in agreement with the BCGs being more massive within the larger aperture region.
A comparison to observations of the values for the whole \textsc{Three Hundred} sample (not only BCGs but all galaxies) can be seen in \citet{Cui18}. For \gx, the code we are using in this work, good agreement is found in general, with the colours being a bit lower than observations ($\sim 0.1 - 0.2$). However, since in this work we are mainly interested in relative changes, these small differences are not relevant for us.

Comparing the absolute magnitudes in \Fig{fig:cmdz0} to observational studies of BCGs like \citet{Pipino2011} we see that our values are significantly larger. This deviation can be due to different reasons. First, due to the nature of our sample, which is a mass complete sample of the largest clusters in a full 1$\hGpc$ side box, at $z=0$. The cluster sample in \citet{Pipino2011} is 85 per cent complete, and spans over redshifts 0.1 to 0.3, so that our magnitudes can be expected to be higher. Even when taking this into account, the differences with observations are still significant. However, this issue is not only seen in \textsc{The Three Hundred} simulations but also in other simulations. For instance, in \citet{Bottrell2017} they find that galaxies in the Illustris simulation \citep{Vogelsberger2014a} are roughly twice as large and 0.7 mag brighter on average than galaxies in the SDSS. In \citet{Vogelsberger2014a} it is also shown that the number of galaxies at the bright end of the galaxy luminosity function is higher than in observations. Regarding BCGs, both IllustrisTNG \citep{Springel2018,Pillepich2018} and C-Eagle \citep{Barnes2017,Bahe2017} simulations possess significantly more massive BCGs than observed \citep{Bahe2017,Henden2020}. This effect can cause the magnitudes to be higher too, as seen in \Fig{fig:cmdz0}. In contrast to this, \citet{Ragone-Figueroa2018} obtain BCG masses from cosmological simulations that are in good agreement with observations, ascribing these results to a better control on the SMBH centering in the host galaxy.

\subsubsection{Merger-induced Colour-Magnitude Changes}
Given the colour-magnitude values, we are interested in how they are affected by mergers. For a qualitative analysis of this, we can first compare the colour-magnitude diagrams of the merger and the control sample. \Fig{fig:cmds} shows this comparison (black dots symbolise the merger sample and blue diamonds the control one) at the two relevant times before and after the merger, only for $R_{ca}=50$ kpc. Above each plot and to its right, the distribution of each variable is shown (solid line for merger sample and dash-dotted for control), together with their median values. We can see that the situation before the merger is quite different to the situation after it. While before the merger (left panel) both samples are indistinguishable, in the right panel one sample is moved with respect to the other, which is confirmed by the distinct distributions in the smaller panels. The BCGs in the merger sample become, in general, brighter and, although both samples increase their colour (i.e. they become redder), we can see that the increase is significantly smaller for the merger sample.
Although we restricted the presentation of the results in \Fig{fig:cmds} to the $R_{ca} = 50$ kpc BCGs for clarity, the conclusions are similar for the other aperture values, and we will discuss them in more detail below.

The results in \Fig{fig:cmds} are in agreement with our previous findings in \Sec{sec:results_stellarmass}, BCGs grow in stellar mass with a certain fraction of it being newly formed (and hence blue) stars. This entails that they become more luminous and do not redden as quickly as their counterparts from the control sample. This can be even better quantified by directly studying the change in these properties.

In general, given a certain cluster property, a simple way of quantifying its change between two points $z_1$ and $z_2$ is using the `fractional difference', computed as:
\beq
\mathrm{FD}=-\frac{P(z_1)-P(z_2)}{P(z_1)},
\label{eq:fdif}
\eeq
where $P(z_{i})$ simply means the value of said property labelled as $P$ at redshift $z_i$. As described before, for our merger study we have two interesting intervals to work with, and so the pair $(z_1, z_2)$ can be $(z_\mathrm{start}, z_\mathrm{end})$ and $(z_\mathrm{before}, z_\mathrm{after})$, respectively. In the first case, since the cluster is not relaxed yet at $z_\mathrm{end}$, we study the immediate effects of mergers and how clusters are affected until the moment the merger is completed. The second case is more relevant for long-term effects, since it allows us to study the consequences of mergers that last even when the cluster has dynamically recovered from the merger.

 \begin{figure*}
  \hspace*{-0.1cm}
  \includegraphics[width=15cm]{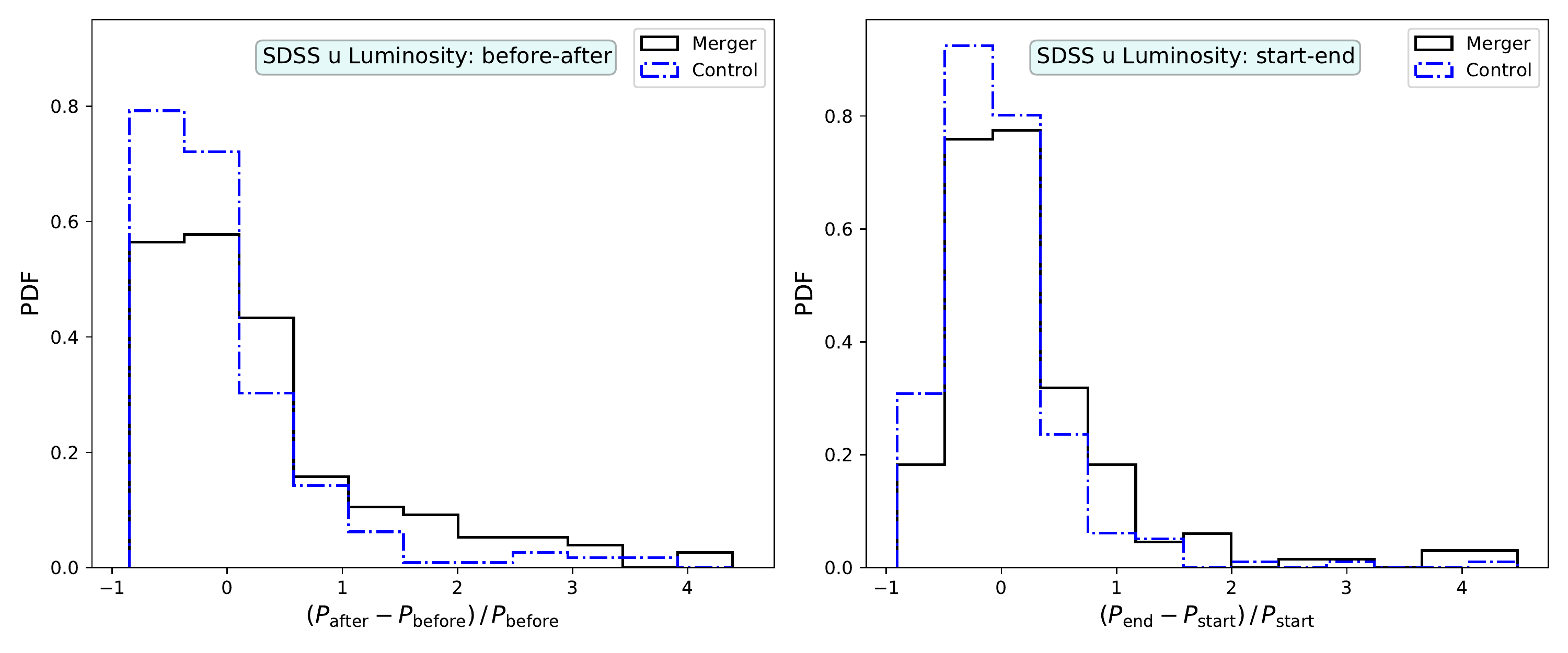}
  \caption{Distributions of fractional difference for SDSS$-u$ luminosity obtained using \Eq{eq:fdif} with $z_1=z_{\mathrm{before}}$ and $z_2=z_\mathrm{after}$ (left) and $z_1=z_{\mathrm{start}}$ and $z_2=z_\mathrm{end}$ (left). Results for the BCGs in the merger sample (solid black) and the control sample (dash-dotted blue), using $R_{ca}=50$ kpc.}
\label{fig:FdifsLum}
\end{figure*}

 \begin{figure*}
  \hspace*{-0.1cm}
  \includegraphics[width=15cm]{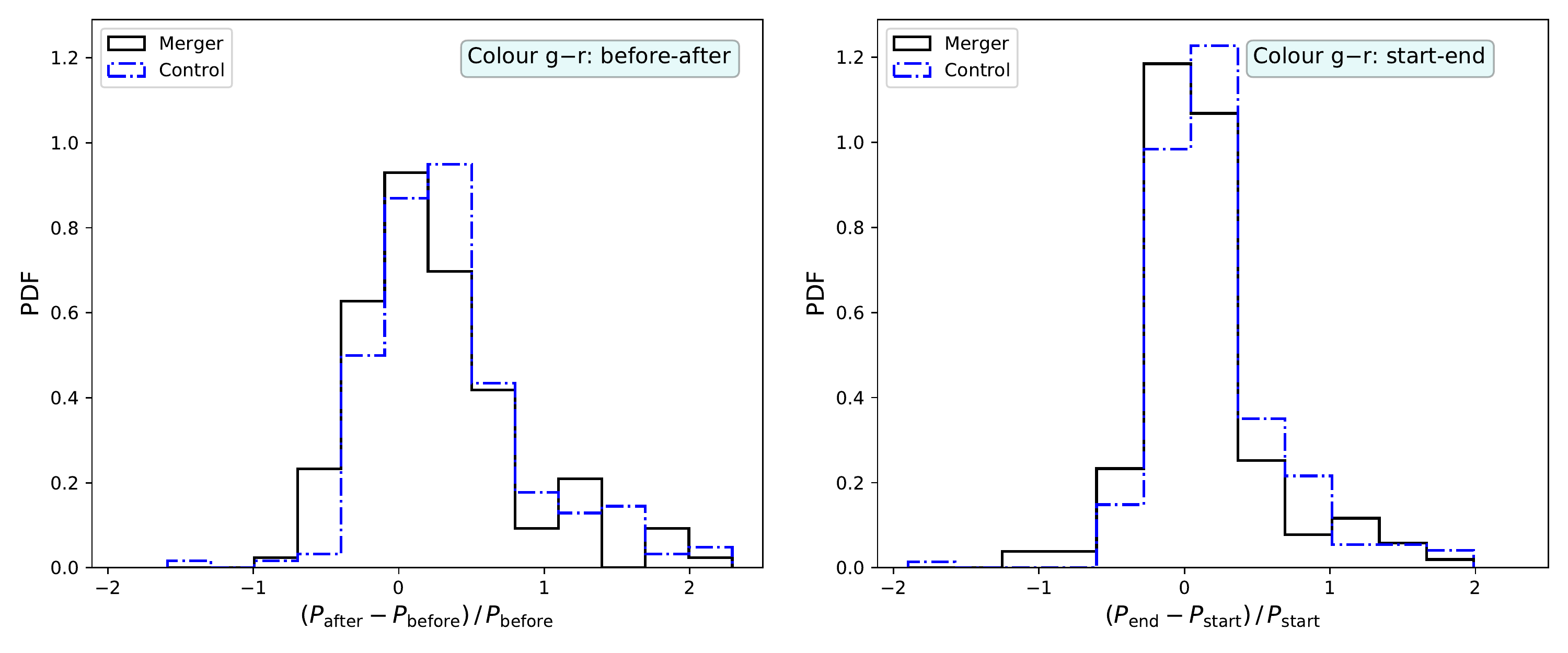}
  \caption{Distributions of fractional difference for colour index $g-r$ obtained using \Eq{eq:fdif} with $z_1=z_{\mathrm{before}}$ and $z_2=z_\mathrm{after}$ (left) and $z_1=z_{\mathrm{start}}$ and $z_2=z_\mathrm{end}$ (left). Results for the BCGs in the merger sample (solid black) and the control sample (dash-dotted blue), using $R_{ca}=50$ kpc}
\label{fig:FdifsColor}
\end{figure*}

In \Fig{fig:FdifsLum} we show the distribution of the fractional difference obtained using \Eq{eq:fdif} for the luminosity of the BCGs in the SDSS-$u$ band, which we chose in order to focus on the young hot stars. In each plot we compare the distributions obtained for the merger sample (solid black line) and control sample (dash-dotted blue line). Again, in the plot we only show the results for $R_{ca}=50$ kpc for clarity, but we will discuss the other values below. We first focus on the left panel, which studies the whole merger phase between $z_{\rm before}$ and $z_{\rm after}$. We see that, for the control sample the distribution peaks at negative values, meaning that for most BCGs the SDSS-$u$ luminosity is decreasing during the merger. On the other hand, looking at the merger sample, it can be seen that the distribution does not show such a prominent peak, and it is moved to higher values of the fractional difference, showing that decreases in luminosity are less frequent in this sample. This effect is less remarkable when checking the right panel, which compares only changes taking place between the start and end of the merger itself. Here both distributions peak at around 0, meaning no significant change in luminosity. While we quantify this even more below (also including the other apertures), here we can already draw some conclusions. As seen for the control sample, BCGs tend to decrease their luminosity in the SDSS-$u$ band in the time interval studied. This can be attributed to the stars getting older. On the contrary, newly formed stars are bright at these wavelengths, and hence the decrease in luminosity is smaller for the merger sample, where a certain amount of new stars is being formed (see \Fig{fig:new_stars}), adding to the SDSS-$u$ luminosity. 

\Fig{fig:FdifsColor} shows the same results as \Fig{fig:FdifsLum} but for the colour index $g-r$. For this index we have to keep in mind that an increase in its value means that the actual colour is becoming redder and vice versa. In this case \Fig{fig:FdifsColor} shows that, when looking at the whole merger phase (left panel) the difference is not as significant as with luminosity. However, we still see that the merger sample is slightly shifted to the left, meaning that BCGs in the control sample are becoming redder than those in the merger sample. In the right panel, that focuses only on the interval $[z_{\rm start},z_{\rm end}]$, the difference between the two distributions is even smaller. 

 \begin{figure*}
  \hspace*{-0.1cm}
  \includegraphics[width=13cm]{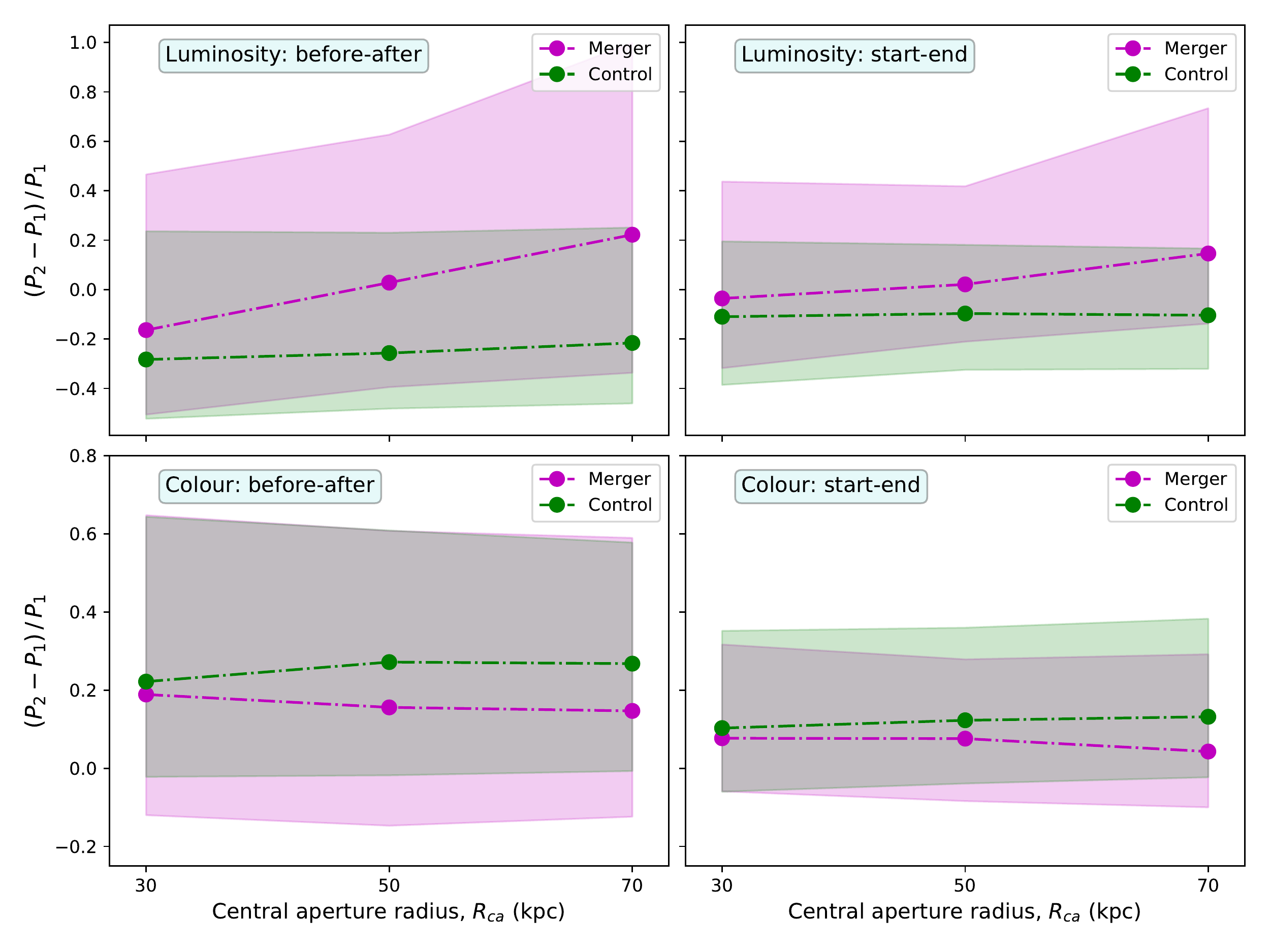}
  \caption{Median values (dots) and 25 and 75 percentiles (shaded regions) of the distributions of the fractional difference computed for two different properties and time periods. All plots compare the results of the merger (magenta) and the control (green) samples. The plots in the upper row are for SDSS-$u$ luminosity, and those in the lower row are for $g-r$ colour index. For both rows, the plots on the left are for the whole merger phase ($z_\mathrm{before}$ to $z_\mathrm{after}$), while those on the right compare the time of the merger ($z_\mathrm{start}$ to $z_\mathrm{end}$).}
\label{fig:FdifsMedians}
\end{figure*}

\begin{table}
\centering
\caption{Resulting $p$-values of the K-S test conducted between the distributions of the indicated fractional differences for the merger and the control sample.}
\begin{tabular}{ccccc} 
\hline
\multicolumn{1}{c}{} & \multicolumn{2}{c}{\textbf{Lum SDSS-}$\bm{u}$} & \multicolumn{2}{c}{\textbf{Colour} $\bm{g-r}$}  \\ \hline \hline
\multicolumn{1}{c|}{$\bm{R_{ca}}$ \textbf{(kpc)}}  & {Before-After}  & {Start-End}   & {Before-After}  & {Start-End}\\ \hline
\textbf{30}            & 0.302              & 0.133        & 0.345       & 0.800     \\
\textbf{50}            & $<10^{-3}$    & 0.015        & 0.295       & 0.256     \\ 
\textbf{70}            & $<10^{-5}$          & $<10^{-6}$    & 0.098       & 0.005     \\ \hline
\end{tabular}
\label{table:FdifsKS}
\end{table}

For a quantitative assessment of the similarity between the distributions, we have computed the two-sample Kolmogórov-Smirnov (K-S) test between each merger-control sample pair. We have done this not only for the distributions shown for the BCGs defined using $R_{ca}=50$ kpc, but also for the same distributions for 30 and 70 kpc. For $R_{ca}=30$ kpc, we see that for both luminosity and colour the p-values are high, meaning that the two samples in each pair could have been drawn from the same distribution. For the luminosity within $R_{ca}=50$ kpc, the low p-value for the whole merger phase confirms that mergers do make a difference in BCGs' luminosity, an effect that is intensified for 70 kpc. This situation does not remain for the colour, where the p-values are still high for 50 kpc, and hence we cannot say that the two distributions are actually distinct. The changes become more significant for 70 kpc, where the p-values decrease. This can also be explained by star formation, since we have seen that more stars are formed in this region than in the others.

To better quantify all of these findings, we have computed the median values of each of the distributions in Figs.~\ref{fig:FdifsLum} and \ref{fig:FdifsColor}, as well as the first and third quartiles. We have done this for the three different regions defining BCGs, and we show all these values in \Fig{fig:FdifsMedians}. The results for the merger sample (in magenta) are compared with those for the control sample (in green). The dots depict the median value of each distribution, while the shaded regions are the 25 and 75 percentiles. This figure allows for an easy comparison of the effects of mergers on luminosity and colour of the involved BCGs. The upper row of \Fig{fig:FdifsMedians} shows the results for the SDSS-$u$ luminosity. For the whole merger phase (left panel), we see that, regardless of the region considered, the BCGs in the control sample decrease their luminosity in $\sim$20 per cent in median during this time. On the other hand, for BCGs undergoing mergers, the luminosity increases up to $\sim$20 per cent for the outermost region of 70 kpc, while the difference is significantly smaller for $R_{ca}=30$ kpc. Similar trends hold when looking at the shorter time interval between $z_\mathrm{start}$ and $z_\mathrm{end}$ (right panel), although the differences between merger and control sample are now smaller. 
We have further checked the correlation between the change in luminosity and the stellar mass growth (not explicitly shown here though) and found a Spearman correlation coefficient of $\sim$0.40 for the three values of $R_{ca}$. Moreover, when considering only the growth due to stars recently formed (see \Fig{fig:new_stars}) the correlation grows to $\sim$0.45 while it decreases to $\sim$0.3 when considering only the accreted stars. This suggests that star formation produces the increase (or better formulated, reduced decrease) in luminosity in the merger sample, so that for the control sample, where fewer stars are formed, the luminosity decreases, driven by the aging of stars. The lower row of \Fig{fig:FdifsMedians} shows the same results as the upper one, but for the colour index $g-r$. We see again that, for the innermost region, the distributions for the merger and the control sample are essentially the same (as inferred from \Tab{table:FdifsKS}). For $R_{ca}=50$ and 70 kpc we can see that the medians are higher for the control sample. This is in agreement with our previous findings, as seen in \Fig{fig:FdifsColor}, that BCGs in mergers become less red than those growing `normally'. However, it is important to note that, although there is a difference in the medians, the differences are very small for the two time intervals considered. 

To sum up, we find that the luminosity of BCGs in clusters that have recently experienced a major merger is expected to be higher than the luminosity of those that did not. This is due to the fact that mergers produce an increase in star formation in the BCG, with new stars being hot and bright. Although a small difference can already be seen for the innermost region of the BCG, the effects are significant and stronger in the outer regions of these galaxies. BCGs in mergers also accrete a significant number of stars ($\sim$4-6 times more stars than those formed during the merger), and although they contribute to the luminosity increase too, this contribution is less relevant. Regarding the colour, the accretion of older stars combined with the aging of all the stars makes BCGs become redder in both merging and non-merging clusters. This colour increase is a bit slower in the merger sample (due to newly formed stars), although the difference might not be significant enough to be observable for $R_{ca}=30$ and 50 kpc. The effect is stronger for the $R_{ca}=70$ kpc region, which is also where star formation is the highest.

\section{Conclusions} \label{sec:conclusions}

In this work we studied the highly energetic events of galaxy cluster mergers as they happen in cosmological simulations. These events are very important in the hierarchical model of structure formation, where they play an important role in the growth of structure. Regarding the multiple effects they can have on both the cluster itself and its individual galaxies, we focused on the brightest cluster galaxies, which include the most massive and luminous galaxies in the universe. We studied how cluster mergers affect their stellar component, and how this reflects on their observable properties colour and luminosity. 

We used the sample of 324 numerically modelled galaxy cluster haloes provided by \textsc{The Three Hundred} project, all with $M_{200} \geq 6.4 \cdot 10^{14} \hMsun$. We tracked the evolution of the central haloes found at $z=0$ back to the highest redshift where they could be found. Using their mass accretion history, we identified possible merger situations as mass growths of at least 100 per cent happening in less than half a dynamical time of the cluster. The relaxation parameter $\chi_\mathrm{DS}$ was used to quantify the dynamical state of clusters. By studying its evolution around the time of the merger as identified in the MAH, we distinguished different characteristic merger times: these are given by the times when the clusters are relaxed (i.e. reach a peak in their $\chi_\mathrm{DS}$ curve) before and after the merger, and the time when they are most disturbed (i.e. reach a minimum) after a merger (see \Fig{fig:DeltaMDS} for a depiction). As we were interested in the effect of mergers on internal cluster properties, we discarded, for resolution reasons, objects with $N < 10000$ at the largest redshift found (which in median means a cut at $z \sim 3.4$). We also removed objects for which the mass growth was too high, $\Delta M / M > 3$; such instances are reminiscent of mis-identifications in the merger tree \citep{Srisawat13}. We ended up with a list of 178 objects. Ab initio, they  simply define a sample of `fast mass growing' clusters, but -- as verified in \Fig{fig:MassRatio21} which shows the ratio between the two most massive merging components -- they can be associated with actual major mergers. Regarding their timescales, while mergers themselves take less than one dynamical time $t_d$ (as imposed by our definition), the whole merger phase lasts in median $3\times t_d$ (although the values can be as large as $7-8\times t_d$). Having a well-defined merger sample, we also created a control sample imitating its features, but with no significant mass growth of the haloes.

Defining the BCG alternatively as the region within 30, 50 and 70 kpc of the halo centre, we compared the stellar content of BCGs before and after the merger. By doing this for both the merger and the control sample, we found that BCGs in mergers accrete around twice as much stellar mass as those just `normally' growing. Comparing the stellar mass formed during the merger we found, using radii 50 and 70 kpc, that BCGs in mergers also form around 70 per cent more stars than those in the control sample. This number drops to 30 per cent when considering only the innermost region of 30 kpc. 

It could be argued that BCGs are not strictly always located at the halo centre, a situation that is in general related to galaxy clusters mergers. However, given the way our applied halo finder \ahf\ works, we can be confident that this is not affecting our results. \ahf\ locates the halo centre as a peak in the density field, and hence $R_{ca}$ will always be placed at the highest density peak, which is precisely the BCG. 
Besides, previous studies \citep{Martel2014,DePropis2021} have shown that, although BCGs might not be in the centre when their host cluster is unrelaxed, they always go back to the centre after some relaxation time (see also \citealp{DeLuca2021} for a similar study using \textsc{The Three Hundred} dataset).  And since for this analysis we have only used the characteristic times $z_\mathrm{before}$ and $z_\mathrm{after}$, which are the times when the cluster is most relaxed, we can be confident with our method. 
For this same reason, we are not worried about possible `halo swaps' due to halo finder problems during a merger \citep[see][]{Behroozi15}, given that we focus our study on the times \textit{before} and \textit{after} the merger.

We analysed the mass-weighted age-metallicity distributions for all the stellar particles in the BCGs, as these are the properties that enter into the calculation of galaxies' luminosities and magnitudes via \textsc{stardust}. We did this for the merger and control samples before and after the merger (\Fig{fig:2dhists}). We found that, although the differences were small, there was a slight increase of young stars after the merger in the merger sample with respect to the control sample, which we attributed to the newly formed stars previously found. 

To assess the effect of mergers on readily accessible observational properties, we studied the luminosity in the SDSS-$u$ band and $g-r$ colour of the involved BCGs (computed with \textsc{stardust}), again for the three different central aperture radii. For the BCGs within 70 kpc we found that, along the whole merger phase, the luminosity of BCGs in clusters that underwent a merger increased by $\sim$20 per cent in median. This was opposed to a decrease of the same amount in the control sample. For the 50 kpc region we also found an increase in luminosity but only of 5 per cent in median for the merger sample, opposed to a 20 per cent decrease in the control sample, while the difference is even smaller using a 30 kpc radius. We attribute this change in luminosity to the new stars that are being formed during the merger, which are more numerous in the outer regions. It has been shown \citep{Perez-Gonzalez2003} that even a few very young stars can make a difference in galaxies' observable properties. There is also a less relevant contribution from the accreted stars, whose combined mass is also higher in the outer regions (see \Fig{fig:new_stars}).

Regarding the colours, we computed the colour index $g-r$ and compared its values at the four characteristic times. We found that BCGs in both the merger and control samples become redder with time in general, which is attributed to the aging of their stellar component. However, we found that this trend is slightly slowed down by cluster mergers. Again, the more significant effects occur within 70 kpc of the halo centre, where the colour growth is 14 per cent slower in the merger sample. Similar differences were found for the smaller regions within 50 and 30 kpc, although in this case they are not statistically significant, and thus might not be observed. These results are in agreement with our previous results regarding stellar mass changes, confirming that there is a burst in star formation in the BCG due to the cluster merging, which is stronger in the outer regions of the BCG (up to 70 kpc), and which is not the main mechanism for the galaxy to grow.

Previous studies have shown that galaxy cluster mergers can produce a burst in star formation in the individual galaxies, which in turn can change their observed colours \citep{Roettiger96,Miller2005,Johnston-Hollitt2008}. These studies do not focus on BCGs, but on the whole galaxy population of the cluster. In respect to BCGs, some works, like \citet{Pipino2009} and \citet{Cerulo2019}, have investigated blue BCGs, relating them to recent star formation. However, the relation between star formation in BCGs and major cluster mergers has not been that thoroughly studied in the literature. Even so, we believe it is relevant for understanding both the evolution of BCGs and the importance of major cluster mergers in the cosmological context. Nevertheless, in future studies we could also include these individual galaxies for a more complete analysis.

\section*{Acknowledgements}
\addcontentsline{toc}{section}{Acknowledgements}

We thank the anonymous referee for raising interesting questions whose answers helped to improve the paper.

This work has been made possible by \textsc{The Three Hundred} (\url{https://the300-project.org}) collaboration. The simulations used in this paper have been performed in the MareNostrum Supercomputer at the Barcelona Supercomputing Center, thanks to CPU time granted by the Red Espa\~{n}ola de Supercomputaci\'on. As part of \textsc{The Three Hundred} project, this work has received financial support from the European Union’s Horizon 2020 Research and Innovation programme under the Marie Sklodowskaw-Curie grant agreement number 734374, the LACEGAL project.

AC, AK, RM and GY are supported by the MICIU/FEDER through grant number PGC2018-094975-C21. AK further thanks Prefab Sprout for Steve McQueen. RH acknowledges support from STFC through a studentship. WC is supported by the European Research Council under grant number 670193 and by the STFC AGP Grant ST/V000594/1. He further acknowledges the science research grants from the China Manned Space Project with NO. CMS-CSST-2021-A01 and CMS-CSST-2021-B01.MDP acknowledges support from Sapienza University in Rome thanks to Progetti di Ricerca Medi 2019 and 2020, RM11916B7540DD8D and RM120172B32D5BE2. SEN is member of the Carrera del Investigador Cient\'{\i}fico of CONICET. He acknowledges support by the Agencia Nacional de Promoci\'on Cient\'{\i}fica y Tecnol\'ogica under the grant PICT-201-0667. 

\section*{Data Availability}
The results shown in this work use data from \textsc{The Three Hundred} galaxy clusters sample. These data are available on request following the guidelines of \textsc{The Three Hundred} collaboration, at \url{https://www.the300-project.org}. The data specifically shown in this paper will be shared upon request to the authors.



\clearpage
\bibliographystyle{mnras}
\bibliography{archive}

\appendix


\bsp	
\label{lastpage}
\end{document}